\documentclass[12pt,a4paper]{article}
\pdfoutput=1
\usepackage{jheppub,tcolorbox}
\usepackage[utf8]{inputenc}
\usepackage{amsmath,amssymb}
\usepackage{float}
\usepackage{tikz} 
\usetikzlibrary{decorations.markings} 
\usetikzlibrary{decorations.text} 
\usetikzlibrary{positioning} 
\usetikzlibrary{backgrounds} 
\usetikzlibrary{fit} 
\usetikzlibrary{calc} 
\usetikzlibrary{shapes} 

\tikzset{dot/.style={draw,circle,inner sep=.7pt,fill,node
    distance=1cm}} 
\tikzset{dot1/.style={draw,circle,inner sep=.7pt,fill}} 
\tikzset{triangle/.style={draw,regular polygon, regular polygon
    sides=3}} 
\tikzset{->-/.style={decoration={
  markings,
  mark=at position .5 with {\arrow{>}}},postaction={decorate}}} 
\tikzset{-<-/.style={decoration={ 
  markings,
  mark=at position .5 with {\arrow{<}}},postaction={decorate}}}
\newcommand\be{\begin{equation}}
\newcommand\ee{\end{equation}}
\newcommand\bea{\begin{eqnarray}}
\newcommand\eea{\end{eqnarray}}

\begin{document}

\begin{titlepage}
\renewcommand{\thefootnote}{\fnsymbol{footnote}}

\vspace*{1.0cm}

\begin{center}
{\textbf{\huge Schwarzschild black hole states and entropies on a nice slice
}}
\end{center}
\vspace{1.0cm}

\centerline{
\textsc{\large J. A.  Rosabal}
\footnote{j.alejandro.rosabal@gmail.com}
}

\vspace{0.6cm}

\begin{center}
\it Asia Pacific Center for Theoretical Physics, Postech, Pohang 37673, Korea
\end{center}

\vspace*{1cm}

\centerline{\bf Abstract}

\begin{centerline}
\noindent

In this  work, we define a quantum gravity  state on a nice slice. The nice slices provide a foliation of spacetime and avoid regions of strong curvature. We explore the topology and the geometry of the manifold  obtained from a nice slice after evolving it in complex time. We compute its associated semiclassical thermodynamics entropy for a 4d Schwarzschild black hole. Despite the state one can define on a nice slice is not a global pure  state, remarkably,  we get a similar result to Hawking's calculation. In the end, we discuss the entanglement entropy of two segments on a nice slice and comment on the relation of this work with the replica wormhole calculation.

\end{centerline}
\thispagestyle{empty}
\end{titlepage}

\setcounter{footnote}{0}

\tableofcontents

\newpage

\section{Introduction}\label{sec:1}

Any attempt to describe the black hole (BH) evaporation using a low-energy effective description, such as semiclassical quantum gravity, must be formulated on the nice slices \cite{Lowe:1995ac, Mathur:2009hf, Polchinski:2016hrw}. These are Cauchy surfaces that foliate spacetime. On these slices, the high energy degrees of freedom decouple from the low energy ones, and thus the low energy effective description does not break down.

Nowadays, a common question asked in the literature associated to  BH evaporation is, where is the mistake in the Hawking's original derivation \cite{Hawking:1974sw, Hawking:1976ra}?. If we can call it a mistake, which perhaps  is  too strong an asseveration, his mistake was not to use such a slicing to specify the quantum gravity (QG) state of the BH.

The starting point of a quantum calculation is the definition of a quantum state. In QG, a state  (here we focus  on the  state produced by complex time evolution)  can be defined on any three-surface embedded in the four-dimensional spacetime \cite{Hartle:1983ai}.
The nice slicing of  the Kruskal spacetime  allows us to define QG  state on a particular nice slice and perform some semiclassical calculations.

Although the existence of these surfaces have been implicitly assumed in some works \footnote{Quoting \cite{Lowe:1995ac}: {\it While it is seldom spelled out, the existence of such a set of surfaces is implicitly assumed in much of the existing literature on black hole evaporation.}}, neither a definition of QG state on them has been presented, nor a calculation of its associated entropies assuming the existence of these slices explicitly,  exists in the literature.

At this point, we find it appropriate to clarify that by QG state, we mean the state of the geometry combined with the state of the matter fields. There have been several remarkable and inspiring works studying only the state of the radiation on the (fixed) BH geometry using the nice slice foliation, in the context of quantum field theory on curved space \cite{Giddings:2020dpb,Giddings:2017mym,Giddings:2012bm,Giddings:2007ie,Giddings:2006be}.

In this work, we define a new QG  state   on the  Kruskal spacetime, particularly  for a Schwarzschild BH,  on a nice slice following the ideas of Hartle and Hawking \cite{Hawking:1983hj, Hartle:1983ai} for the wave function of the universe. Then, using this state, we compute its associated entropies. Rather Remarkably, we get similar results as of that in \cite{Gibbons:1976ue, Hawking:1978jz} for the thermodynamic entropy. Nevertheless, the main and more striking difference with  \cite{Gibbons:1976ue, Hawking:1978jz} is that on these slices, it is impossible to define a global pure  state for BH's. A direct consequence of this impossibility is that we can not use a wave function to describe this  state. Instead, we must use a density matrix to describe the global mixed state on a nice slice, in the same spirit of \cite{Hawking:1986vj} and \cite{Page:1986vw}.

Although we do not include matter in this first proposal, we leave windows open to include it in future works. The advantage of this calculation is that we can trust it until a very late time when studying BH evaporation.

It is worth to remark that apparently there is a big problem with the nice slices when considering them in studying BH evaporation; see \cite{Polchinski:2016hrw} for a discussion about it. The issue is that in order to avoid the region of strong curvature (close to the singularity), to keep the effective description valid, the interior portion of all nice slices must be fixed at the  Schwarzschild coordinate $\text{r}_0<2\text{M}$, (see the red line in  Fig. \ref{Nice_Slice}). This portion only grows in time. As it grows in time, the number of bits on it coming from the Hawking's pairs in the radiation also grows indefinitely, leading to a linear dependence in time for the entropy of the radiation $S_{rad}(t)\sim \text{cons}\times t$. We refer to \cite{Hartman:2013qma} and references therein for a discussion on the time-dependent entropy of a BH.

This unpleasant fact conflicts with the unitary evolution of quantum mechanics. Quantum mechanically, the entropy should grow until some time $t_p$, called Page time, and then decrease to zero when the evaporation is completed, following the so-called Page curve \cite{Page:1993wv, Page:2013dx}.

One might see this problem as an obstruction to use the nice slices in this setup; however, it is not. Recently there have been remarkable proposals where this problem can be overcome; for AdS space in two dimensions \cite{Almheiri:2019qdq, Penington:2019kki}, and for asymptotically flat space  \cite{Gautason:2020tmk, Anegawa:2020ezn,Hashimoto:2020cas,Hartman:2020swn}. In these works, the nice slicing of a BH has been implicitly assumed too.

The paper is organized as follows. In section \ref{sec:2}, we review the construction of state  produced by complex time evolution, density matrix, and partition function in QG. Then we exemplify these constructions presenting the Hawking's calculation of the Schwarzschild BH thermodynamic entropy. In section \ref{sec:2}, we introduce the concept of nice slice and define the  state on a particular one. We explore the complex sections' topology and geometry defined by evolving a nice slice in complex time. This complex manifold defines a semiclassical global mixed state. Using it, we compute its associated thermodynamic entropy. In section \ref{sec:5}, we introduce the density matrix interpretation of this state. After this discussion, in section \ref{sec:6} we point out the relation of our work with \cite{Almheiri:2019qdq, Penington:2019kki, Gautason:2020tmk, Anegawa:2020ezn, Hashimoto:2020cas, Hartman:2020swn}, and we make some remarks on the entanglement entropy and replica wormholes on a nice slice. Conclusions are presented in section \ref{sec:7}.

\subsection{Summary of the results}

This paper explores the QG groud state defined on a nice slice  embedded in the Kruskal spacetime;  for a Schwarzschild BH. It is defined by complex time evolution. The nice slice where the state is defined can be placed anywhere in the Kruskal spacetime, even overlapping the horizons. When a nice slide overlaps the horizons, we can reach null infinity and perform some semiclassical calculations in this region. It is the region where we can compute, for instance, the time-dependent entanglement entropy for an evaporating BH.

The new geometry we present here does not correspond to a semiclassical global pure state. It is a global mixed state whose description is supplied by a density matrix
$\rho\big[h_{ij}^{+},\phi_0^{+}; h^{-}_{ij},\phi^{-}_0 \big]$, where $(h_{ij}^{+},\phi_0^{+})$ and $( h^{-}_{ij},\phi^{-}_0)$ are the values of the three-metric and the matter fields on the boundaries of the complex-extended manifold. These boundaries correspond to the slice of the Lorentzian space where the state is defined.

When evolving a nice slice in complex time, we find that the metric on it is complex, and the topology of the complex-extended manifold resembles a cylinder. In other words, the boundaries of the density matrix are connected by a surface. This fact supports that the state one can define on any nice slice is a global mixed state. Higher genus topologies can be considered too; however, we do not explore those geometries here.

Remarkably, the semiclassical state described by this geometry leads to a thermodynamic entropy which corresponds to the expected one for a two-sided BH, despite the state is not a  global pure state, in contrast to the Hartle Hawking state. We have followed similar steps to those in the original Hawking's derivation for performing all the calculations.

The complex time evolution of a nice slice is not straightforward; the main reason for this is that a portion of the nice slice remains fixed inside the horizon. This portion grows in Lorentzian time but does not evolve forward. For this portion, the complex extension is driven only by the metric's boundary values on the boundaries of the portions that explicitly depend on time.

One of the exciting features this geometry presents is that it intersects the Lorentzian space in two surfaces. This feature allows us to split the complex-extended manifold in two manifolds. Each of these manifolds have a density matrix associated, and they can be regarded as the building blocks of the original density matrix. In other words, the density matrix factorizes as
\be
\rho\big[h_{ij}^{+},\phi_0^{+}; h^{-}_{ij},\phi^{-}_0 \big]=
\int\text{D}h_{ij}^{1}\text{D}\phi_0^{1}\rho_{+}\big[h_{ij}^{+},\phi_0^{+}; h^{1}_{ij},\phi^{1}_0 \big]\rho_{-}\big[h_{ij}^{1},\phi_0^{1}; h^{-}_{ij},\phi^{-}_0 \big].
\ee
Each of these manifolds represent semiclassical amplitudes from a surface in the past to a future surface. These two surfaces can overlap the past and future horizon, in which case the building blocks of $\rho\big[h_{ij}^{+},\phi_0^{+}; h^{-}_{ij},\phi^{-}_0 \big]$ can be regarded as $S$ matrices.

This paper also discusses the entanglement entropy associated to the semiclassical QG  state described above. We explain how the replica manifold must be built. Due to some ambiguities in extending the nice slice's portion that does not evolve forward in Lorentzian time, we find that the density matrix associated with the replicated manifold contains contributions from the disconnected as well as connected geometries. The concept of replica wormhole naturally arises in this setup.

Following a similar logic to that in the construction of the density matrix of the universe \cite{Hawking:1986vj, Page:1986vw}, we construct the most general density matrix we can associate to the replicated manifolds. It is given by
\be
\tilde{\rho}(n)=\underset{\text{disconnected}}{ \tilde{\rho}^n} +\underset{\text{connected}}{ \tilde{\rho}_n}\label{mre_den_m},
\ee
where $n$ is the number of replicas.
Within this logic, we argue how, from the definition \eqref{mre_den_m}, the so-called factorization problem \cite{Penington:2019kki} can be avoided and, hence, no ensemble average is needed to make the setup consistent. In the end, we argue on how this construction must be extended to account for the proper definition of information flux at null infinity.

\section{State, density matrix and partition function}\label{sec:2}

In QFT, the state of a system can be specified by giving its wave functional if the state is pure or its associated density matrix if the state is mixed. A state of interest in QFT is the ground state. It can be defined by a path integral \cite{Hartle:1983ai}, with boundary on a given spacelike three-surface $\Sigma$ of spacetime labeled by some time,  $t=t_0$. The four-manifold that defines the state can be obtained by evolving  in complex time the three-surface $\Sigma$. Equivalently it can be simply stated  as  $t_0 \rightarrow t_0 -\text{i}\tau$. After choosing some boundary conditions on the boundary of the four-manifold,  the state can be written as
\be
\Psi[\phi_0(x),t_0]=\int \text{D}\phi(x,\tau)\text{exp}\big(\text{i}\text{I}[\phi(x,\tau)]\big),
\ee
where $\text{I}[\phi(x,t)]$, is the action of the system. The wave functional $\Psi[\phi_0(x),t_0]$, gives the amplitude that a particular field configuration $\phi_0(x)$, happens to be on the spacelike surface $t=t_0$. The path integral is over all fields for $\tau<0$, which match $\phi_0(x)$, on the surface $\tau=0$, (the $\tau=0$, surface corresponds to the $t=t_0$, slice of spacetime).
Having the state on the $t_0$-slice one can evolve the state to a different $t$-slice. Formally it can be stated as
\be
\Psi[\phi_0(x),t]=\text{exp}\big[-\text{i}(t-t_0)\hat{\text{H}}\big] \Psi[\phi_0(x),t_0],\label{evol}
\ee
where $\hat{\text{H}}$ is the Hamiltonian operator of the system. Expression \eqref{evol} can be regarded as a formal solution of a Schr{\"o}dinger-like equation for the wave functional $\Psi[\phi_0(x),t]$,
\be
\text{i}\partial_{t}\Psi[\phi_0(x),t]=\hat{\text{H}}\Psi[\phi_0(x),t],\label{schrodingerQFT}
\ee
with initial conditions $\Psi[\phi_0(x),t_0]$.

In QG, as there is no well-defined measure of the location of a particular spacelike surface in spacetime, the state's definition differs from that in QFT.  Despite this fact, a  state can be defined  by complex time evolution. Following \cite{Hartle:1983ai} one can define a wave functional for a  state of a gravitational system as
\be
\Psi[h_{ij}, \phi_0]=\int \text{D}g\text{D}\phi \text{exp}\big(\text{i}\text{I}[g,\phi]\big).\label{qg_state}
\ee
Where, now $\text{I}[g,\phi]$, is the gravitational and the matter action defined over a complex section of the original space \footnote{The usual prescription in QG, as in QFT, is to set a foliation of the space labeled by some time $t$, and then pick a particular slice $\Sigma$ which corresponds to $t=t_0$, and evolve it in imaginary time, i.e., $t_0\rightarrow t_0-\text{i} \tau$, see for instance \cite{hartman_lec}.}. The integration, in this case, is over all matter fields and the four-geometries which match $\phi_0(x)$, and the induced three-metric $h_{ij}$, on a boundary $\Sigma$ that belongs to the real space (Lorentzian manifold). Up to this point, we consider that $\Sigma$ divides the Lorentzian manifold in two parts.

Another important quantity in QFT and QG is the probability $P[h_{ij},\phi_0]$  that a particular field configuration occurs on  $\Sigma$.  When the state is pure  it is defined as
\be
P[h_{ij},\phi_0]=\Psi[h_{ij}, \phi_0]\Psi^{*}[h_{ij}, \phi_0].\label{probabilidad00}
\ee
Combining \eqref{probabilidad00} and \eqref{qg_state} one can regard the  probability as a path integral over the four metrics defined on the manifold resulting from gluing the original four-manifold (with a boundary $\Sigma$) that defines $\Psi[h_{ij}, \phi_0]$, with another copy of itself, albeit with an opposite orientation.
They share the same boundary, and the path integral is obtained by integrating  over the field configurations defined on the resulting manifold which match $(h_{ij},\phi_0)$, on $\Sigma$

The total probability $\text{Z}$ (or the partition function) is given by
\be
\text{Z}=\int\text{D}h_{ij}\text{D}\phi_0 P[h_{ij},\phi_0],\label{total_proba}
\ee
where the integration is over the values of the fields on $\Sigma$.
From \eqref{total_proba} and \eqref{qg_state} one can see that the total probability is a path integral over the four metrics defined on the manifold resulting from gluing the original four-manifold  with another copy (with opposite orientation) of itself.
\be
\text{Z}=\int \text{D}g\text{D}\phi \text{exp}\big(\text{i}\text{I}[g,\phi]\big).
\ee

In addition it is possible to define a density matrix
\be
\rho\big[ h^{+}_{ij},\phi^{+}_0 ; h_{ij}^{-},\phi_0^{-} \big]=\Psi\big[ h^{+}_{ij},\phi^{+}_0 \big]\Psi^{*}\big[h_{ij}^{-},\phi_0^{-} \big].
\ee
As it clearly factorizes,  it is associated to a pure state. The diagonal elements gives us the probability as in \eqref{probabilidad00},  $P\big[ h_{ij},\phi_0\big]=\rho\big[ h_{ij},\phi_0 ; h_{ij},\phi_0 \big]$; and its trace gives us the partition function, similarly to \eqref{total_proba}, 
\be
\text{Z}=\text{Tr}[\rho]=\int\text{D}h_{ij}\text{D}\phi_0 \rho\big[ h_{ij},\phi_0 ; h_{ij},\phi_0 \big].
\ee

In the previous definition of the density matrix, we first have to perform an integration over two disjoint manifolds. Then, the trace operation glues them together over the surface $\Sigma$.

Soon after this proposal for the gravitational state came out Hawking and Page \cite{Hawking:1986vj, Page:1986vw} realized that in principle one can include contributions from geometries that connect the boundaries of the density matrix. For recent applications of this procedure see \cite{Anous:2020lka, Chen:2020tes}. We shall call them {\it connected geometries}. In this case, the state can not be considered  pure, hence the only object available to describe it would be a density matrix of the form
\be
\rho\big[h^{+}_{ij},\phi^{+}_0 ; h_{ij}^{-},\phi_0^{-} \big]=\sum_{m,n}C_{mn}\Psi_m\big[ h^{+}_{ij},\phi^{+}_0 \big] \Psi^{*}_n\big[h_{ij}^{-},\phi_0^{-} \big],\label{mixed_den}
\ee
where $C_{mn}$ does not factorizes i.e., we can not find a basis where the matrix $C_{mn}$ factorizes as $C_{mn}= c_m c_n$.
In fact, by allowing connected geometries, the boundary $\Sigma$ does not need to divide the real space into two parts.

As more involved geometries are allowed the density matrix can have more than two boundaries on the real space. The trace operation over the non-observable boundaries \footnote{These are the boundaries which do not correspond to the boundary where we are interested in computing the observables \cite{Hawking:1986vj}.} gives rise to what could be regarded as a reduced density matrix with boundary values on the real space, on the remaining boundary. Like for the pure state, a trace over the remaining boundary (observable boundary) gives us the partition function $\text{Z}$ of the system. This can be regarded as a path integral over the disconnected and connected geometries resulting from gluing the boundaries of the manifolds that previusly defined a density matrix with several boundaries \cite{Hawking:1986vj, Page:1986vw}
\begin{align}
\text{Z}=\text{Tr}\big[\rho \big]=\int{D}h_{ij}\int{D}\phi_0\sum_{m,n}C_{mn}\Psi_m\big[ h_{ij},\phi_0 \big] \Psi_n^{*}\big[h_{ij},\phi_0 \big]=\nonumber \\
\sum_{\begin{matrix}\text{disconnected}\\ +\ \text{connected}\end{matrix}}\int \text{D}g\text{D}\phi \text{exp}\big(\text{i}\text{I}[g,\phi]\big).\label{Tr_mixed_den}
\end{align}

We end the discussion of the section with the time evolution in quantum gravity. Any state in QG must be a solution to the Wheeler-Dewitt equation \cite{DeWitt:1967yk}. This equation states that
\be
\hat{\text{H}}\Psi[h_{ij},\phi_0]=0,\label{WDW1}
\ee
or
\be
\hat{\text{H}}\rho[ h^{+}_{ij},\phi_0^{+}; h^{-}_{ij},\phi_0^{-}]=\delta\big[h_{ij}^{+},h_{ij}^{-}\big]\delta\big[\phi_0^{+},\phi_0^{-}\big]\label{WDW2}.
\ee
It is a Schr{\"o}dinger-like equation for the wave functional, or the density matrix of a gravitational system, and $\hat{\text{H}}$ is the gravitational Hamiltonian operator including the matter contribution. This Schr{\"o}dinger-like equation differs enormously from that in QFT \eqref{schrodingerQFT}. Note that time does not appear explicitly in the equations above.

From \eqref{WDW1} or \eqref{WDW2}, it is not difficult to see that the unitary evolution of QFT \eqref{evol} does not apply to quantum gravity. Nevertheless, in the semiclassical approximation, when a metric is fixed and a foliation specified, one can recover a Schr{\"o}dinger-like equation for the state of the matter in that particular metric \cite{Halliwell:1989dy}.

\subsection{Hawking's calculation}\label{sec:Hawking's calculation}

As an example of the previous constructions, we review the Hawking's calculation of the Schwarzschild black hole entropy. For simplicity we focus only on the gravitational contribution  \cite{Gibbons:1976ue, Hawking:1978jz}, which leads to the famous formula $\text{S}_{BH}=\frac{\text{A}}{4}$, where $\text{A}=16\pi\text{M}^2$,  is the area of the horizon.

Let us compute the partition function of a gravitational system in vacuum. Our starting point is the gravitational state
\be
\Psi[h_{ij}^{-}]=\int \text{D}g \text{exp}\big(\text{i}\text{I}[g]\big),\label{start_state}
\ee
where
\be
\text{I}[g]=(16\pi)^{-1}\int\limits_{V} dx^4\sqrt{-g} \ R+(8\pi)^{-1}\int\limits_{\partial V} dx^3\sqrt{-h} \big[K \big],\label{EHaction}
\ee
with
\be
\int\limits_{\partial V} dx^3\sqrt{-h} \big[K \big]=\int\limits_{\partial V} dx^3\sqrt{-h} \big(K -K_0\big).\label{GHYaction}
\ee
The boundary term in \eqref{GHYaction}, also known as the Gibbons–Hawking–York term, plays a crucial role in finding the entropy of a black hole. To define state we must specify on which three-surface $\Sigma$ we want to define it and the asymptotic behaviour at spatial infinity (after the complex extension) of the metrics we are integrating over. For this case we consider those metrics which are asymptotically flat. To specify $\Sigma$ it is convenient but not necesary to specify a folliation of the space.

\begin{figure}[h]
\centering
\includegraphics[width=.6\textwidth]{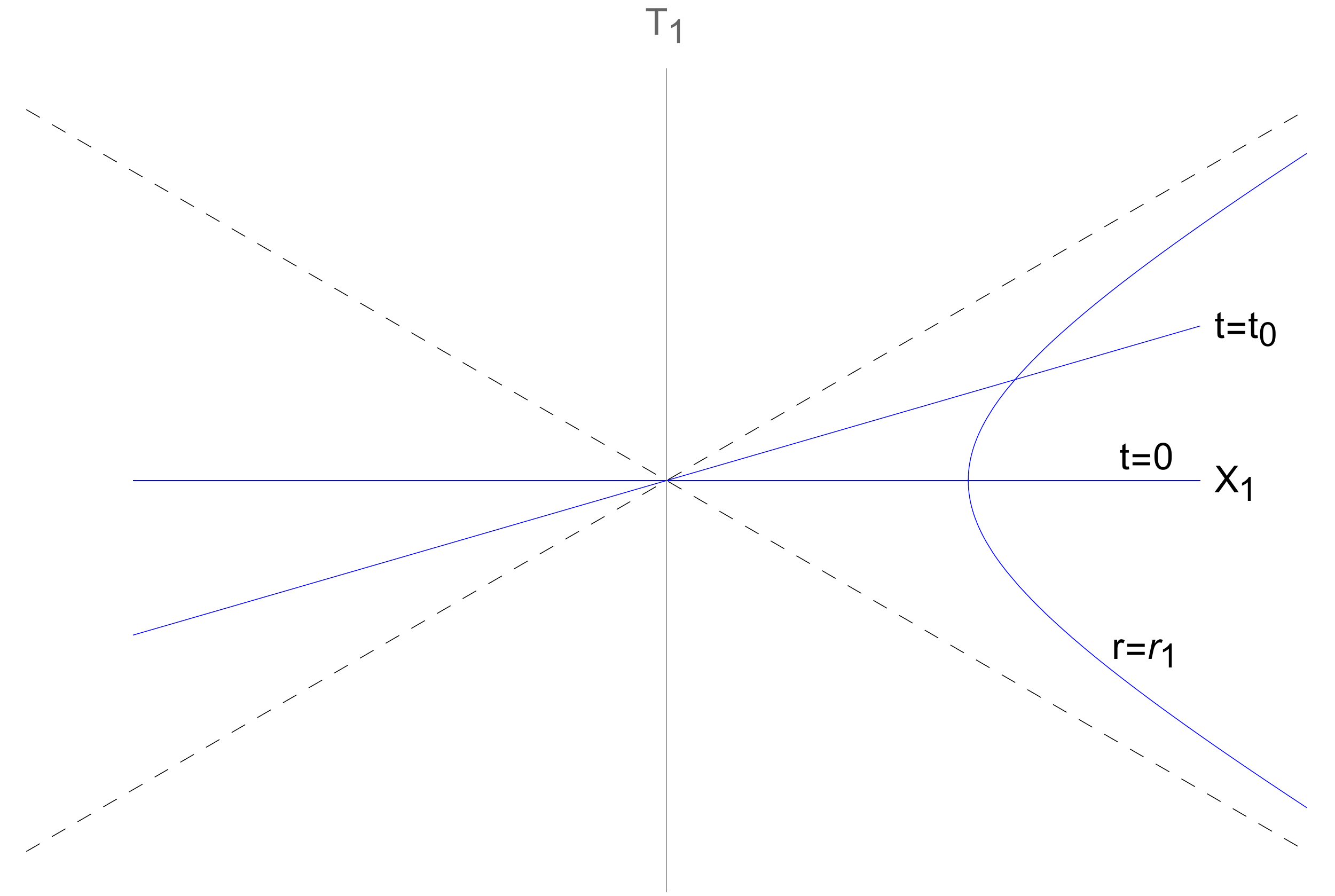}
\caption{\sl Kruskal spacetime and the $(t,\text{r})$ foliation of it.}\label{S_fol}
\end{figure}

In Kruskal coordinates usually we take a folliation that corresponds to an observer at spatial Lorentzian infinity Fig. \ref{S_fol},
\bea\nonumber
\text{T}_1 & = & \ \ \ \sqrt{ (\frac{\text{r}}{2\text{M}}-1)}\text{e}^{\frac{\text{r}}{4\text{M}}}\ \text{sinh}(\frac{t}{4\text{M}}),\\
\text{X}_1 & = & \pm\sqrt{ (\frac{\text{r}}{2\text{M}}-1)}\text{e}^{\frac{\text{r}}{4\text{M}}}\ \text{cosh}(\frac{t}{4\text{M}}).\label{K_coor}
\eea
Coordinates \eqref{K_coor} cover only the left and right wedges in Fig. \ref{S_fol}.
Note that so far, we have not specified the metric of the Kruskal  spacetime, only the foliation. Now we pick a particular spacelike slice. The most popular is $\text{T}_1=0$, ($t=0$), where we can define the so-called Hartle-Hawking (HH) state for a black hole \cite{Hartle:1976tp}. This state is not pure for the portion of space $\text{X}_1\geq 0$. However, it is obtained from the density matrix associated with the global pure state \eqref{start_state} (where we can also include the matter contribution) after tracing over the degrees of freedom on $\text{X}_1<0$.

\begin{figure}[h]
\centering
\includegraphics[width=.5\textwidth]{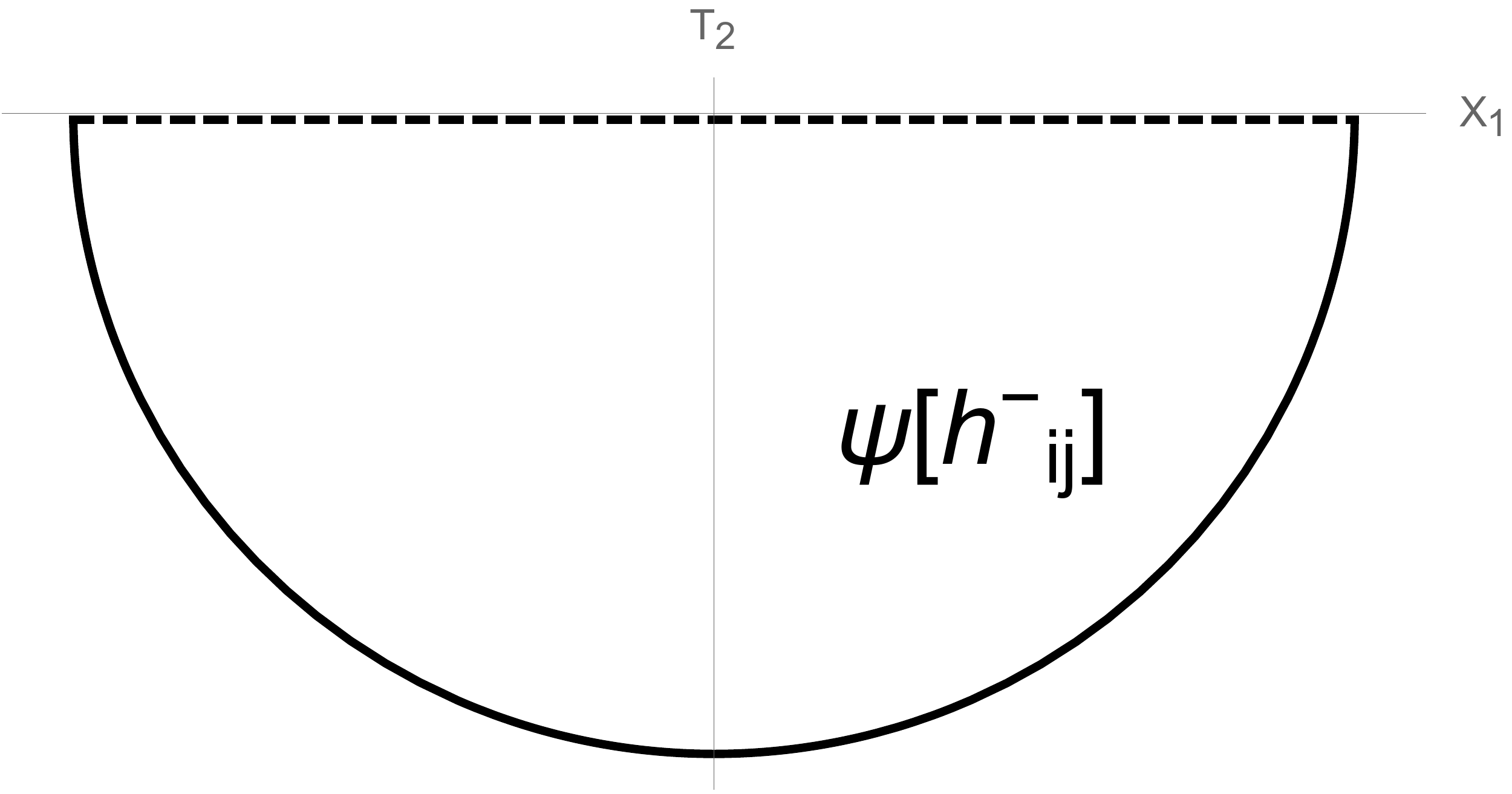}
\caption{\sl Geometric representation of the BH state defined on the $\text{T}_1=0$ slice. }\label{E_state1}
\end{figure}

After the complex extension \footnote{In this case $t_0=0$, which corrsponds with $\text{T}_1=0$.}, $t_0\rightarrow t_0-\text{i}\tau$, $\text{T}_1\rightarrow \text{T}$, and $\text{X}_1\rightarrow \text{X}$,
\bea\nonumber
\text{T} & = & \sqrt{ (\frac{\text{r}}{2\text{M}}-1)}\text{e}^{\frac{\text{r}}{4\text{M}}}\ \text{sinh}(\frac{t_0-\text{i}\tau}{4\text{M}}),\\
\text{X} & = & \sqrt{ (\frac{\text{r}}{2\text{M}}-1)}\text{e}^{\frac{\text{r}}{4\text{M}}}\ \text{cosh}(\frac{t_0-\text{i}\tau}{4\text{M}}). \label{E_coor}
\eea
Where, the periodicity of the $\tau$ direccion, $\tau\sim\tau+8\pi\text{M}$, follows from \eqref{E_coor}.
At least formally, now we can define the state \eqref{start_state}, where $h_{ij}^{-}$, is the boundary value of the path integral on the slice $t_0=0$. This state can be geometrically represented as in Fig. \ref{E_state1}.
Note that we have used $(\text{T}_1,\text{X}_1)$, for the real variables in Kruskal coordinates Fig. \ref{S_fol} and $(\text{T},\text{X})$, for the complex ones of the complexified space \eqref{E_coor}. The $(\text{T}_2,\text{X}_2)$, variables will be reserved only for the imaginary part, for example $\text{T}_2$, in Fig. \ref{E_state1}. The space for the particular choice $t_0=0$, is called the Eucliedan section.
Note also that the axis $\text{X}_1$, ($\text{T}_1=0$, or $\text{T}_2=0$) is common for both, the Lorentzian spacetime and the Euclidean section. In \eqref{E_coor}, we have extended only the right wedge because it is enough to cover the geometry we seek for defining the state \footnote{ At the end of the conclusions, we comment on a possible issue in this procedure\label{speculate}.}

\begin{figure}[h]
\centering
\includegraphics[width=.6\textwidth]{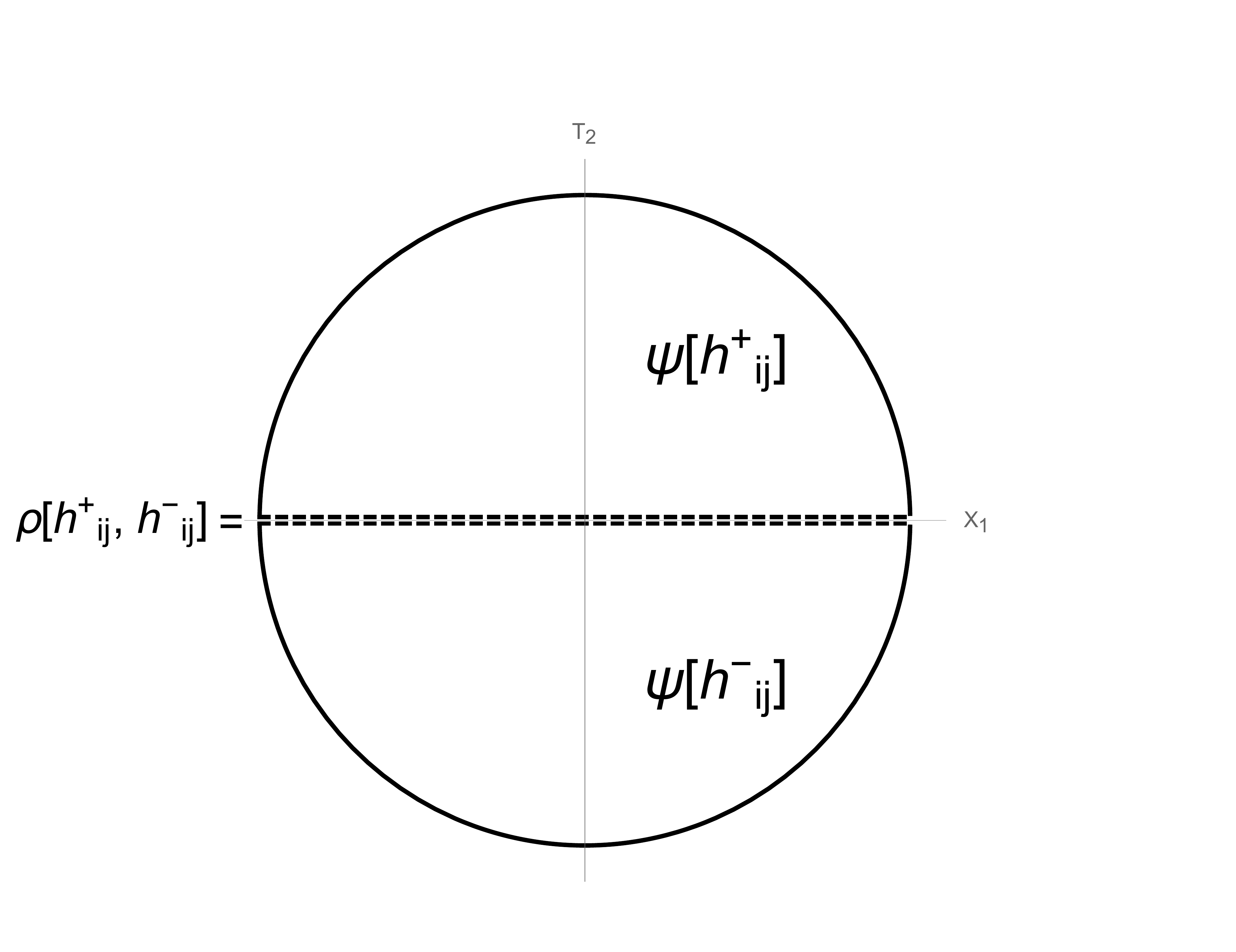}
\caption{\sl The two disjoint geometries that geometrically represent the density matrix associated with the state in Fig. \ref{E_state1}.}\label{E_state2}
\end{figure}

\begin{figure}[h]
\centering
\includegraphics[width=.6\textwidth]{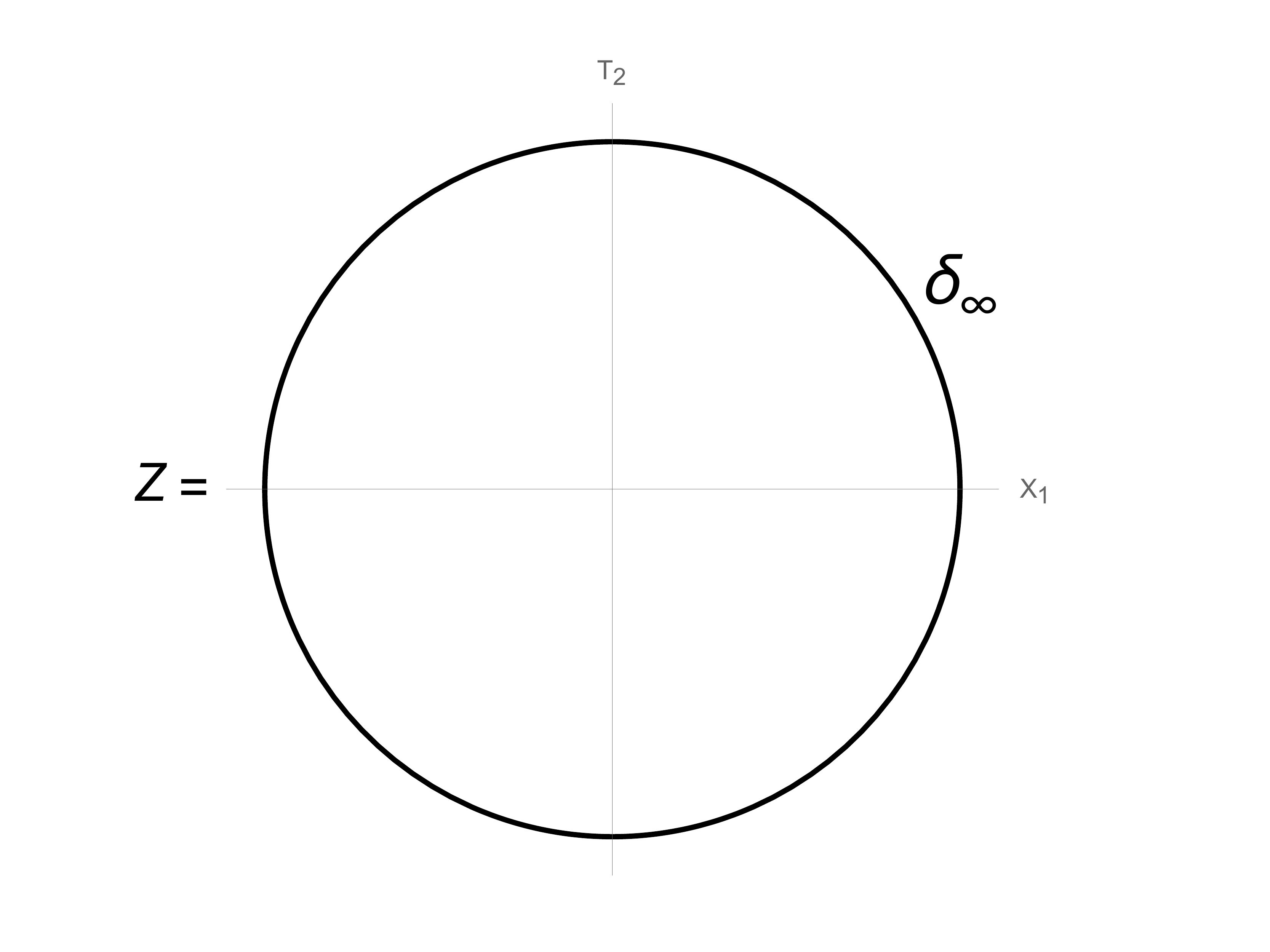}
\caption{\sl The disk geometry that geometrically represents the partition function associated with the state in Fig. \ref{E_state1}.}\label{E_state3}
\end{figure}

Having defined the state we are in a condition of computing the partition function. For that we define the density matrix $\rho[h_{ij}^{+},h_{ij}^{-}]=\Psi[h_{ij}^{+}] \Psi[h_{ij}^{-}]$ (we do not take the complex conjugate because in this case the wave functional is real). It is represented geometrically in Fig. \ref{E_state2}.
This density matrix factorizes in two wave functionals and it is defined through a path integral over two disjoint geometries, which means the state is pure.
The partition function of the system is given by
\be
\text{Z}=\text{Tr}[\rho]=\int{D}h_{ij}\rho[h_{ij},h_{ij}]. \label{Zexample}
\ee
Geometrically the trace operation on the density matrix amounts to gluing the two semi-disk in Fig. \ref{E_state2}. The partition function can be geometrically represented as in Fig. \ref{E_state3},
where $\delta_{\infty}$ represents the boundary of the disk geometry. In the end of the calculation we send it to infinity.

As discussed above, after combining \eqref{Zexample} and \eqref{start_state}, it is not difficult to see that the path integration in the partition function \eqref{Zexample} is over the metrics defined on the resulting manifold in Fig. \ref{E_state3} with flat boundary conditions at spatial infinity.

In the semiclassical aproximation we just evaluate the path integral on a classical solution $g_c$, extracted from the  Einstein's equations,
\be
\text{Z}=\text{Tr}[\rho]=\int{D}h_{ij}\rho[h_{ij},h_{ij}]=\int \text{D}g \text{exp}\big(\text{i}\text{I}[g]\big)\sim \text{exp}\big(\text{i}\text{I}[g_{c}]\big).
\ee
At this point is where we fix the metric by solving the Einstein's equation on the disk Fig. \ref{E_state3}. In the $(\text{T}_2,\text{X}_1)$, coordinates, the vacuum solution takes the form
\be
\text{ds}^2=\frac{32\text{M}^3}{\text{r}}\text{e}^{-\frac{\text{r}}{2\text{M}}}\big(\text{dT}_2^2+\text{dX}_1^2\big)+\text{r}^2\text{d}\Omega^2,\label{E_Krus}
\ee
\be
\text{r}=2\text{M}\big(1+\text{W}_0(\frac{\text{X}_1^2+ \text{T}_2^2}{\text{e}}) \big).
\ee
Which is the Wick rotated version of the Kruskal metric with $\text{r}\geq2\text{M}$, and singularity free. It is convenient to express the metric in the $(\tau,\text{r})$, coordinates \eqref{E_coor}
\be
\text{ds}^2=(1-\frac{2\text{M}}{\text{r}})\text{d}\tau^2+\frac{\text{dr}^2}{(1-\frac{2\text{M}}{\text{r}})}+\text{r}^2\text{d}\Omega^2,
\ee
where $\tau\sim\tau+8\pi\text{M}$. The peridodicity $\beta=8\pi\text{M}$, of the $\tau$ direcction indicates the semiclassical state that is described by this geometry is thermal with a temperature $T=\beta^{-1}$.

In the Euclidean section these two set of coordinates $(\text{T}_2,\text{X}_1)$, and $(\tau,\text{r})$, cover the same space which correspods to the whole disk geometry. Since it is a vacuum solution $R_{\mu\nu}=0$. The only contribution to the action comes from the boundary term. It is given by
\be
\text{I}[g_c](\beta)=(8\pi)^{-1}\int\limits_{\text{r}=\text{r}_{\infty}\rightarrow\infty} dx^3\sqrt{-h} \big[K \big]=4\pi\text{i}\text{M}^2=\text{i}\ \frac{\beta^2}{16\pi},\label{b_contri}
\ee
then
\be
\text{Z}(\beta)=\text{exp}\big[- \frac{\beta^2}{16\pi}\big]\label{b_contri1}.
\ee
Now, using $\text{S}_{BH}=\big(1-\beta \partial_{\beta}\big) \text{ln}\text{Z}(\beta)\Big{|}_{\beta=8\pi\text{M}}$, see  \cite{Gibbons:1976ue, Hawking:1978jz}, \eqref{b_contri1} leads to the famous relation
\be
\text{S}_{BH}=\frac{\text{A}}{4}\ .
\ee

Instead of choosing the slice $t_0=0$, one could have chosen $t_0\neq 0$. For this case, \eqref{E_coor} would be complex. One might see this fact as an obstruction for choosing other slices to define state; however, as we will see in the next section, it is not. In the case under discussion ($t_0\neq0$), it is not needed to perform any further calculation if we want to find the partition function on a different $t_0$-slice. Using only the rotation symmetry of the metric \eqref{E_Krus} (boost symmetry for the Kruskal metric on the real space), one concludes that the partition function and the entropy are invariant under time translations. Although, the state might differ from the one defined on the slice $t_0=0$.

\section{Nice slicing of a Schwarzschild black hole and complex extension}\label{sec:3}

In this section, we shall introduce the concept of nice slices \cite{Lowe:1995ac, Mathur:2009hf, Polchinski:2016hrw}. On these slices, the semiclassical QG calculations for an evaporating black hole do not break down until a very late time. We shall also define the QG  state on a nice slice. Then, using it, we will compute its associated partition function and thermodynamic entropy.

The nice slices are a set of Cauchy surfaces which foliate spacetime. The surfaces avoid regions of strong spacetime curvature (close to singularities) but
cut through the infalling matter and the outgoing Hawking radiation. Importantly, infalling matter and the outgoing Hawking radiation should have low energy in the local coordinates on each slice. We also require that the slices be  smooth everywhere, with small extrinsic curvature compared to
any microscopic scale. With these requirements, we ensure that the effective QG description does not break down, and using this foliation, we can follow the evaporation of a black hole until a very late time.

\begin{figure}[h]
\centering
\includegraphics[width=.6\textwidth]{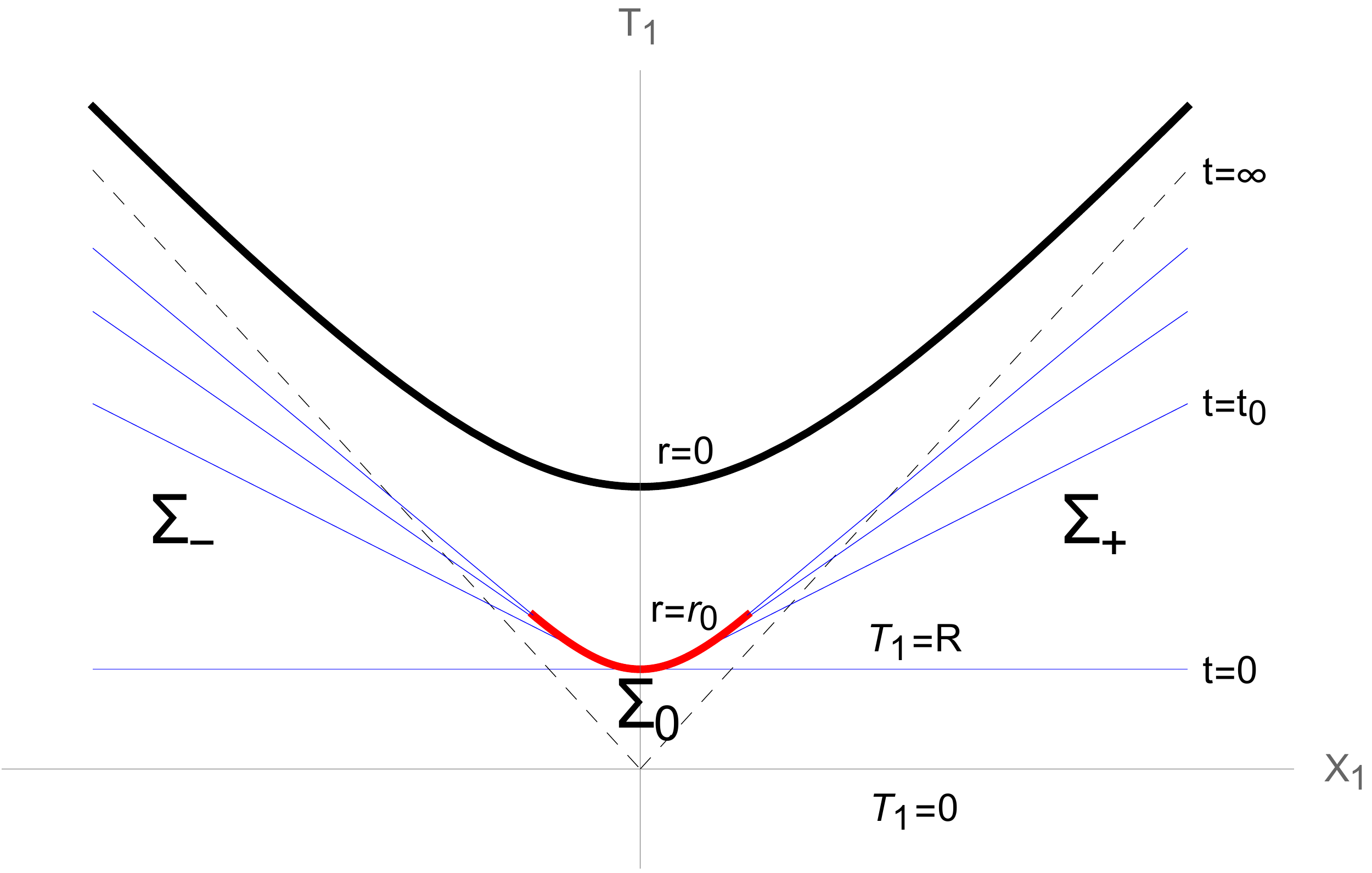}
\caption{\sl Nice slice foliation of a Schwarzschild BH in Kruskal coordinates.}\label{Nice_Slice}
\end{figure}

Conveniently,  one can chose slices that agree with slices of constant Schwarzschild time in the asymptotic region.
A particular set of nice slices is depicted in Fig. \ref{Nice_Slice}. In Kruskal coordinates we use Schwarzschild time $t$ to parameterize them,

\begin{align} \nonumber
\Sigma_{-}: \quad & \text{cosh}(\frac{t}{4\text{M}})\ \text{T}_1+ \text{sinh}(\frac{t}{4\text{M}})\ \text{X}_1 =\text{R} & ; & \quad \text{X}_1<-\text{R}\ \text{sinh}(\frac{t}{4\text{M}}),\\ \nonumber
\Sigma_{0}: \quad &\text{X}_1^2- \text{T}_1^2 =-\text{R}^2 & ; & \quad -\text{R}\ \text{sinh}(\frac{t}{4\text{M}})<\text{X}_1< -\text{R}\ \text{sinh}(\frac{t}{4\text{M}}), \\ \nonumber
\Sigma_{+}: \quad & \text{cosh}(\frac{t}{4\text{M}})\ \text{T}_1- \text{sinh}(\frac{t}{4\text{M}})\ \text{X}_1 =\text{R} & ; & \quad \text{X}_1>\text{R}\ \text{sinh}(\frac{t}{4\text{M}}).
\end{align}
The constant $\text{R}$ is assumed to be large by comparison with any microscopic scale, but small enough to keep the slices far from the singularity. Note that $\Sigma_0$ (the red line in Fig. \ref{Nice_Slice}) only grows as we evolve forward in Schwarzschild time but it is fixed at a constant $\text{r}_0<2\text{M}$.

The metric in Kruskal coordinates of a Schwarzschild BH is
\be
\text{ds}^2=\frac{32\text{M}^3}{\text{r}}\text{e}^{-\frac{\text{r}}{2\text{M}}}\big(-\text{dT}_1^2+\text{dX}_1^2\big)+\text{r}^2\text{d}\Omega^2,\label{L_section_metric}
\ee
where
\be
\text{r}=2\text{M}\big(1+\text{W}_0(\frac{\text{X}_1^2- \text{T}_1^2}{\text{e}}) \big),\label{lambertW0}
\ee
with $\text{W}_0$ the Lambert function.

As in the usual  folliation \eqref{K_coor} of the Schwarzschild space, in the nice slice foliation we can change from the coordinates $(\text{T}_1,\text{X}_1)$, to the coordinates $(t,\text{r})$. For changing coordinates, for example, on $\text{X}_1<0$ we use the relations
\begin{align}\label{soq}
\text{X}_1^2- \text{T}_1^2 = (\frac{\text{r}}{2\text{M}}-1)\text{e}^{\frac{\text{r}}{2\text{M}}},\\ \nonumber
\Sigma_{-}: \ \text{cosh}(\frac{t}{4\text{M}})\ \text{T}_1+ \text{sinh}(\frac{t}{4\text{M}})\ \text{X}_1 =\text{R}.
\end{align}
In the first line of \eqref{soq} we have inverted the relation \eqref{lambertW0}.
The solution of this system of equations is given by
\bea\nonumber
\text{T}_1 & = & +\rho\ \text{sinh}(\frac{t}{4\text{M}}) +\text{R}\ \text{cosh}(\frac{t}{4\text{M}}),\\ \nonumber
\text{X}_1 & = & -\rho\ \text{cosh}(\frac{t}{4\text{M}}) -\text{R}\ \text{sinh}(\frac{t}{4\text{M}}),
\eea
with
\be
\rho=\sqrt{ (\frac{\text{r}}{2\text{M}}-1)\text{e}^{\frac{\text{r}}{2\text{M}}}+\text{R}^2}.\label{rho_definition}
\ee

In the $(t,\text{r})$, coordinates the metric takes the form
\be
\text{ds}^2=-(1-\frac{2\text{M}}{\text{r}})\text{d}t^2+\frac{2\text{R}}{\sqrt{ (\frac{\text{r}}{2\text{M}}-1)\text{e}^{\frac{\text{r}}{2\text{M}}}+\text{R}^2}}\text{d}t\text{dr}+\frac{\text{dr}^2}{1-\frac{2\text{M}}{\text{r}}(1-\text{R}^2\text{e}^{-\frac{\text{r}}{2\text{M}}})}+\text{r}^2\text{d}\Omega^2\label{metricrtcoor}.
\ee
Here, $0\leq t<\infty$, and $\text{r}_{0}<\text{r}<\infty$, where $\text{r}_{0}$, in a solution to the equation $\text{X}_1^2- \text{T}_1^2 = (\frac{\text{r}_0}{2\text{M}}-1)\text{e}^{\frac{\text{r}_0}{2\text{M}}}=-\text{R}^2$, at the boundaries of $\Sigma_{-}$ and $\Sigma_{+}$, i.e.,
\be
1-\frac{2\text{M}}{\text{r}_0}(1-\text{R}^2\text{e}^{-\frac{\text{r}_0}{2\text{M}}})=0\implies \text{r}_{0}=2\text{M}\big(1+\text{W}_0(-\frac{R^2}{\text{e}})\big)<2\text{M}\label{condi_r}.
\ee

\begin{figure}[hbt]
\centering
\includegraphics[width=.6\textwidth]{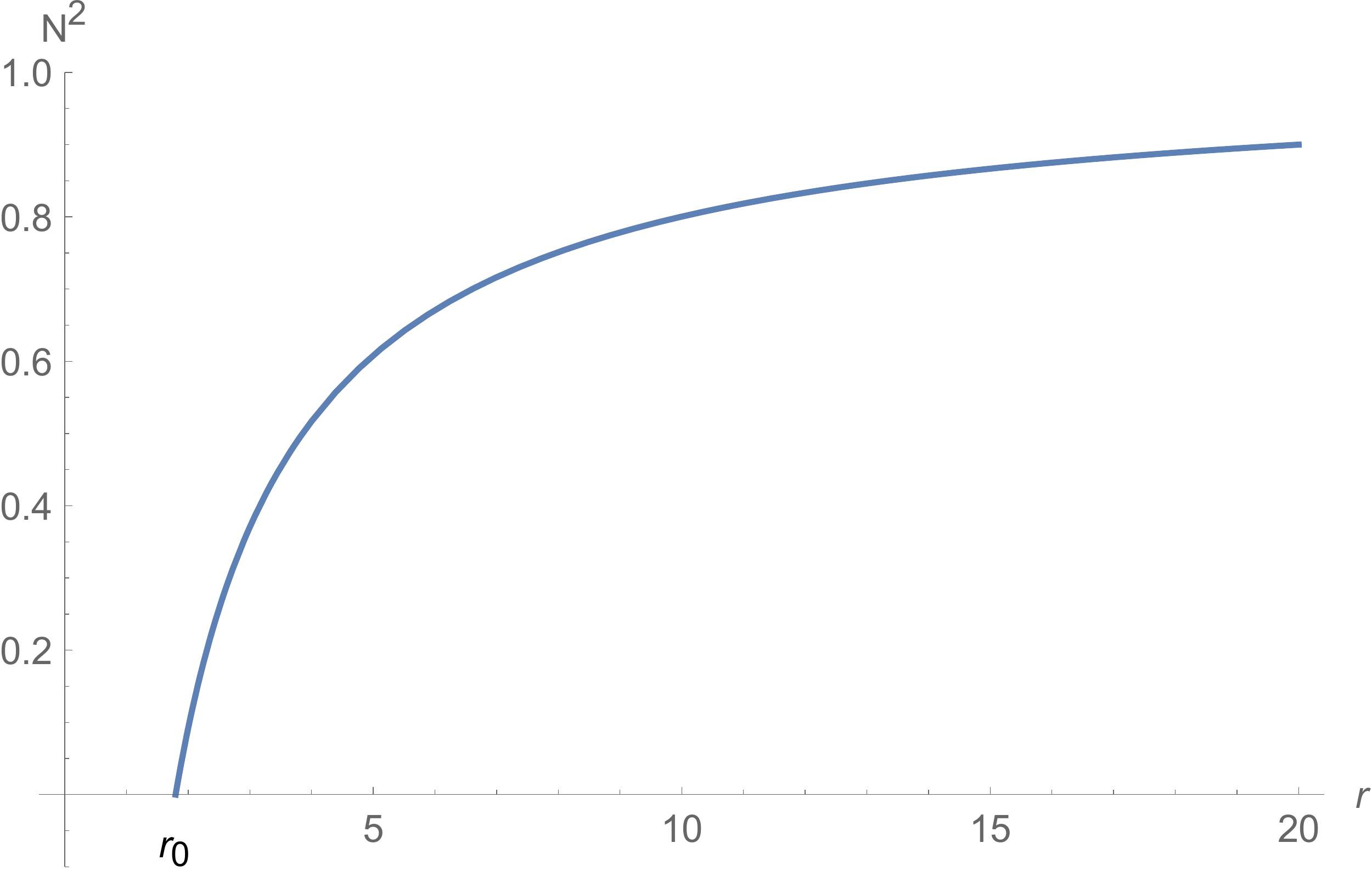}
\caption{\sl Lapse function.}\label{Lapse_Function}
\end{figure}

The metric \eqref{metricrtcoor} can be rewritten in a more suggestive form, making manisfest the foliation and the canonical structure of this geometry
\be
\text{ds}^2=-\text{N}^2\text{d}t^2+\text{h}_{ab}(\text{dx}^a+\text{V}^a\text{d}t)(\text{dx}^b+\text{V}^b\text{d}t) \label{sugges_metric}.
\ee
In the ADM form \eqref{sugges_metric}, \cite{Deser:1959zza, Arnowitt:1959ah, Arnowitt:1960es}, we have: $\text{N}^2=1-\frac{2\text{M}}{\text{r}}(1-\text{R}^2\text{e}^{-\frac{\text{r}}{2\text{M}}})$, $\text{V}^1=\text{R}\big(\frac{2\text{M}}{\text{r}} \big)^{\frac{1}{2}}\text{e}^{-\frac{\text{r}}{4\text{M}}}\text{N}$, $h_{ab}=\text{diag}(\text{N}^{-2},\text{r}^2,\text{r}^2\text{sin}^2(\theta))$, and $\text{dx}^a=(\text{dr},\text{d}\theta,\text{d}\phi)$. Note that the lapse function $\text{N}^2$, is non-negative Fig. \ref{Lapse_Function}, and $\text{N}^2(\text{r}_0)=0$, see equation \eqref{condi_r}.

Like in the Hawking's calculation above,  now we can pick an slice and perform the complex extension. Picking the slice $t=t_0$ and extending it,  $t_0 \rightarrow t_0 -\text{i}\tau$, $\text{T}_1\rightarrow\text{T}$, $\text{X}_1\rightarrow\text{X}$, yields to
\bea\nonumber
\text{T} & = & + \rho\ \text{sinh}(\frac{t_0 -\text{i}\tau}{4\text{M}}) +\text{R}\ \text{cosh}(\frac{t_0 -\text{i}\tau}{4\text{M}}),\\ \label{2dsurface}
\text{X} & = & -\rho\ \text{cosh}(\frac{t_0 -\text{i}\tau}{4\text{M}}) -\text{R}\ \text{sinh}(\frac{t_0 -\text{i}\tau}{4\text{M}}).
\eea
The periodicity of the $\tau$ direcction follows from \eqref{2dsurface}, $\tau\sim \tau+8\pi\text{M}$. If we are going to consider that the geometry we are building describes a semiclassical state, this state would have a temperature $T=\beta^{-1}=\frac{1}{8\pi
\text{M}}$ . Note that $\text{X}^2-\text{T}^2= (\frac{\text{r}}{2\text{M}}-1)\text{e}^{\frac{\text{r}}{2\text{M}}}\in \mathbb{R}$, which implies that $\text{r}\in \mathbb{R}$. Also, that $\tau=0$, corresponds to the $t=t_0$, slice in Fig. \ref{Nice_Slice}. For the sake of generality we want to consider \footnote{Note that when $t_0\rightarrow\infty$, $\Sigma_{-}$, and $\Sigma_{+}$ sit on the horizons,  on null  infinity, i.e., on $\mathcal{I}^{+}$ in the Penrose diagram. } $t_0\neq 0$, in \eqref{2dsurface}.

We can see a clear difference when we compare \eqref{2dsurface} with the Euclidean section of the Schwarzschild space \eqref{E_coor} (recall that in \eqref{E_coor} $t_0=0$). The section defined in \eqref{2dsurface} is complex. Moreover, the state defined by \eqref{2dsurface} does not lead to the HH state. As we will see below, now it is more convenient to define a density matrix associated with a global mixed  state to describe it.

Finally the metric on the complex sections corresponding to the extension of the slices denoted by $\Sigma_{-}$ and $\Sigma_{+}$ is
\be
\text{ds}^2=\frac{32\text{M}^3}{\text{r}}\text{e}^{-\frac{\text{r}}{2\text{M}}}\big(-\text{dT}^2+\text{dX}^2\big)+\text{r}^2\text{d}\Omega^2,\label{metriccomplex}
\ee
with $(\text{T},\text{X})$, defined over the complex surfaces
\bea\nonumber
\text{T} & = & +\big (\rho\ \text{sinh}(\frac{t_0 -\text{i}\tau}{4\text{M}}) +\text{R}\ \text{cosh}(\frac{t_0 -\text{i}\tau}{4\text{M}})\big),\\ \label{complex_surface_ext_defini}
\text{X} & = & \pm \big(\rho\ \text{cosh}(\frac{t_0 -\text{i}\tau}{4\text{M}}) +\text{R}\ \text{sinh}(\frac{t_0 -\text{i}\tau}{4\text{M}})\big),
\eea
where the minus sign in the second line of \eqref{complex_surface_ext_defini} corresponds to the extension of $\Sigma_{-}$, and the plus sign to the extension of $\Sigma_{+}$,
and $
\text{r}=2\text{M}\big(1+\text{W}_0(\frac{\text{X}^2- \text{T}^2}{\text{e}}) \big)$.
In the $ (\tau,\text{r})$ coordinates the metric takes the form
\be
\text{ds}^2=\text{N}^2\text{d}\tau^2+\text{h}_{ab}(\text{dx}^a-\text{i}\text{V}^a\text{d}\tau)(\text{dx}^b-\text{i}\text{V}^b\text{d}\tau)\label{complexmetric},
\ee
which follows directly from \eqref{sugges_metric}.
Note that no subscripts  appear in the differential  forms of the metric \eqref{metriccomplex}. Metrics \eqref{metriccomplex} or \eqref{complexmetric} are complex, however this is not an issue in this kind of calculation. Complex metrics have been explored (used) in several guises \cite{Almheiri:2019qdq, Penington:2019kki, Halliwell:1989dy,Brown:1990fk}.

\section{Topology and geometry of the complex sections }\label{sec:4}

This section shall study the topology and geometry of the manifolds obtained by the complex extension. We shall call them $\delta_{\rho}^{-}$ for the extension of $\Sigma_{-}$, and $\delta_{\rho}^{+}$ for the extension of $\Sigma_{+}$.

Expression \eqref{2dsurface} defines a 2d surface,
$\delta_{\rho}^{-}: \{\text{T}=\text{T}_1+\text{i}\text{T}_2$ , $\text{X}=\text{X}_1+\text{i}\text{X}_2$\}\Big/
\bea \label{curve on CT}
\text{T}_1 & = &+\text{R}_1(\rho)\ \text{cos}(\frac{\tau}{4\text{M}}),\nonumber\\
\text{T}_2 & = & -\text{R}_2(\rho)\ \text{sin}(\frac{\tau}{4\text{M}}),\nonumber\\
\text{X}_1 & = & -\text{R}_2(\rho)\ \text{cos}(\frac{\tau}{4\text{M}}),\nonumber\\
\text{X}_2 & = & +\text{R}_1(\rho)\ \text{sin}(\frac{\tau}{4\text{M}}),
\eea
where
\bea
\text{R}_1(\rho) & = & \rho \ \text{sinh}(\frac{t_0}{4\text{M}})+\text{R}\ \text{cosh}(\frac{t_0}{4\text{M}}), \nonumber \\
\text{R}_1(\rho) & = & \rho\ \text{cosh}(\frac{t_0}{4\text{M}}) +\text{R}\ \text{sinh}(\frac{t_0}{4\text{M}})\ .
\eea
We can think about this 2d surface as embedded in $\mathbb{C}^2$ or $\mathbb{R}^4$. Either way we can see that the surface has the topology of an annulus. For each constant $\rho=\rho_0$, the curve $\delta_{\rho_0}^{-}$ is a circumference on a Clifford torus or on $\text{S}^3$. The surface has two boundaries, one at $\text{r}=\text{r}_0$, $(\rho=0)$, and the other at $\text{r}=\text{r}_{\infty}\rightarrow\infty$, $(\rho\rightarrow\infty)$, note that $\rho(\text{r}_0)=0$.

\subsection{Clifford torus construcction for the complex extension of $\Sigma_{-}$} 

\begin{figure}[h]
\centering
\includegraphics[width=.3\textwidth]{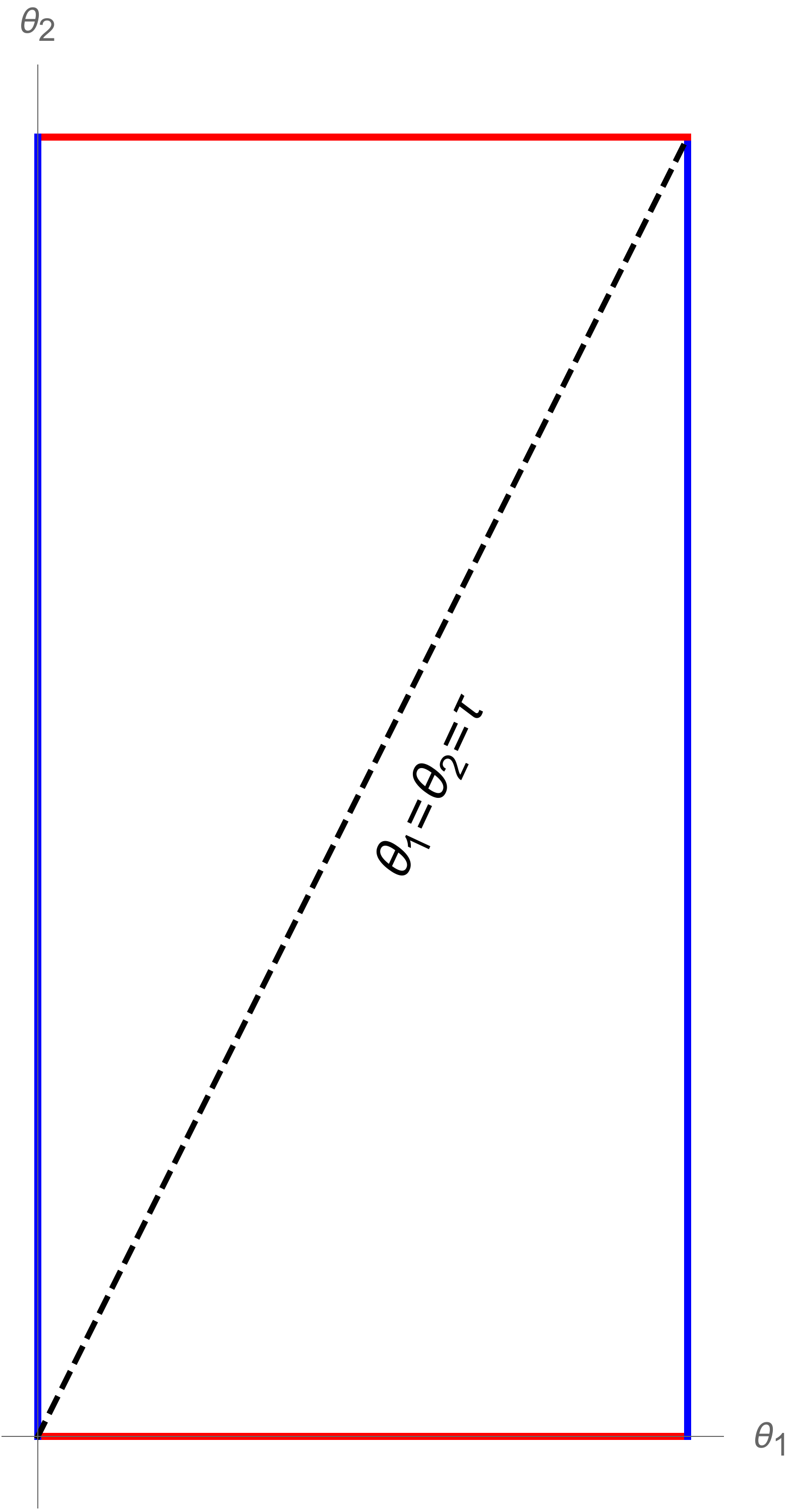}
\caption{\sl Topological representation of the torus. The red and blue lines are identified. The diagonal dashed line represents a circle on the torus.}\label{C_Torus}
\end{figure}

In $\mathbb{R}^4$, for a constant $\rho=\rho_0$, we can define the torus $\textbf{T}^2(\rho_0)=\text{S}^1 \times \text{S}^1:(\text{T}_1,\text{X}_2,\text{T}_2,\text{X}_1)$
\be
=\Big (\text{R}_1(\rho_0)\ \text{cos}(\frac{\theta_1}{4\text{M}}),\text{R}_1(\rho_0)\ \text{sin}(\frac{\theta_1}{4\text{M}}),-\text{R}_2(\rho_0)\ \text{sin}(\frac{\theta_2}{4\text{M}}),-\text{R}_2(\rho_0)\ \text{cos}(\frac{\theta_2}{4\text{M}})\Big).
\ee
Now we pick the curve on the torus, parameterized by $\tau$, $\theta_1=\theta_2=\tau$, we shall call it $\delta_{\rho_0}^{-}$. To see that each curve $\delta_{\rho_0}^{-}$, is an $\text{S}^1$ we use the representation of the torus in Fig. \ref{C_Torus}, where the lines of the same color are idenfied.
From Fig. \ref{C_Torus} it is straightforward to see that the dashed diagonal line is indeed a circle. Finally, joining all the circles $\delta_{\rho_0}^{-}$ from each torus $\textbf{T}^2(\rho_0)$ ($\rho_0$ ranges from zero to infinity), we can easily see that the resulting surface is exactly $\delta_{\rho}^{-}$, \eqref{curve on CT} or \eqref{2dsurface}.

Under similar considerations one can get a 2d surface $\delta_{\rho}^{+}$, from the complex extension of $\Sigma_{+}$
\be
\delta_{\rho}^{+}:\ \Big (\text{R}_1(\rho)\ \text{cos}(\frac{\tau}{4\text{M}}),-\text{R}_1(\rho)\ \text{sin}(\frac{\tau}{4\text{M}}),-\text{R}_2(\rho)\ \text{sin}(\frac{\tau}{4\text{M}}),\text{R}_2(\rho)\ \text{cos}(\frac{\tau}{4\text{M}})\Big).
\ee

So far, we can view this space as two disjoint annulus; or as portions of two disjoint cigar geometries, each one with a boundary at $\text{r}=\text{r}_0$, and the other at infinity, when $t_0\neq0$. When $t_0=0$, these two spaces touch each other at $\text{r}=\text{r}_0$, $(\rho=0)$, on the boundaries
\bea
\delta_{0}^{-}:&{}& \Big(\text{R}\ \text{cos}(\frac{\tau}{4\text{M}}),+\text{R}\ \text{sin}(\frac{\tau}{4\text{M}}),0,0\Big),\nonumber\\
\delta_{0}^{+}:&{}& \Big(\text{R}\ \text{cos}(\frac{\tau}{4\text{M}}),-\text{R}\ \text{sin}(\frac{\tau}{4\text{M}}),0,0\Big)\nonumber.
\eea
Note that $\delta_{0}^{-}\equiv \delta_{0}^{+}$, but they have diffrent orientation.

\begin{figure}[h]
\centering
\includegraphics[width=.7\textwidth]{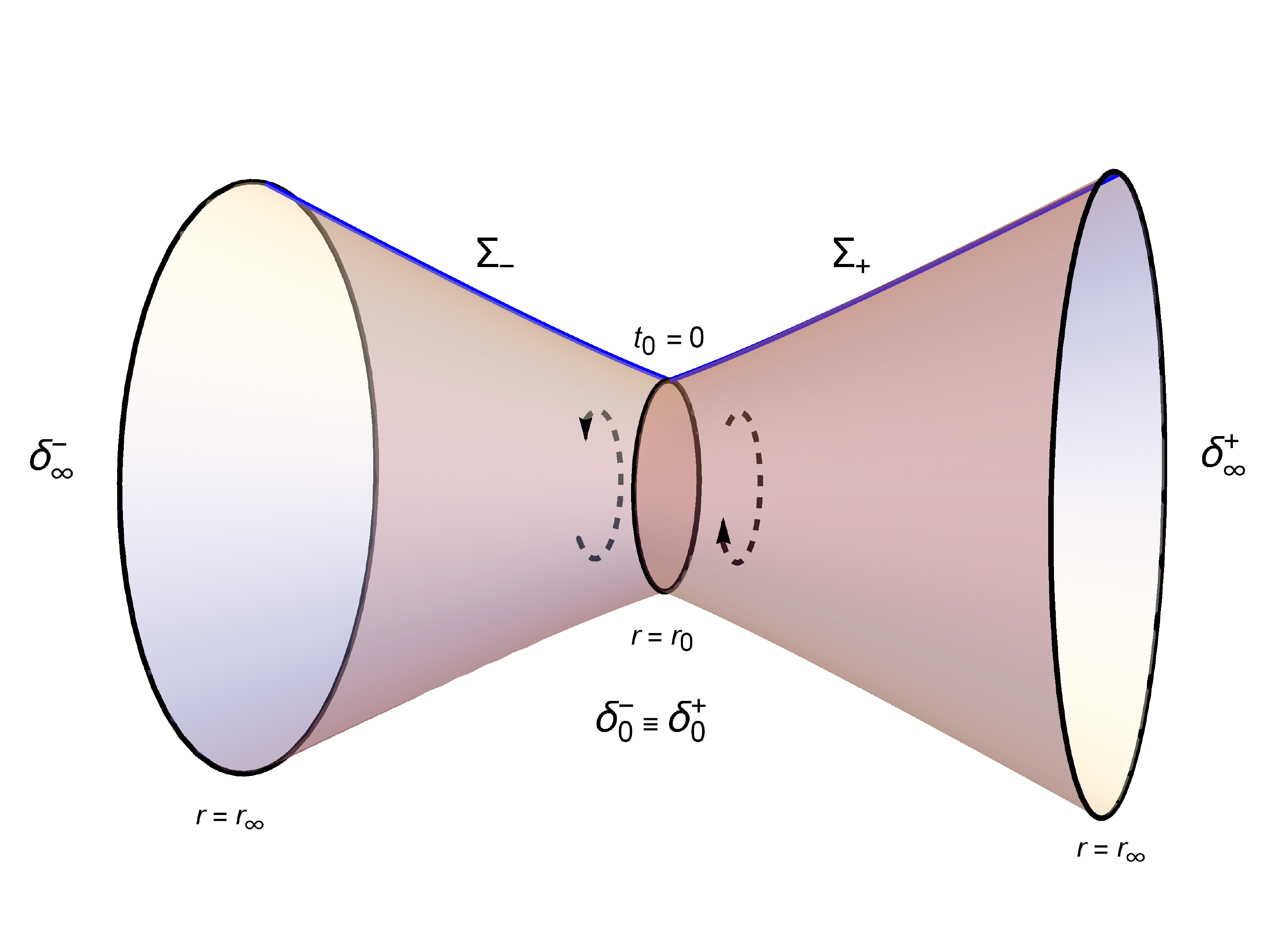}
\caption{\sl Two-dimensional representation for $t_0=0$, of the 2d surfaces $\delta_{\rho}^{-}$ and $\delta_{\rho}^{+}$ emmbeded in four dimensions. No red slice appear in this case.}\label{extension0}
\end{figure}

\begin{figure}[h]
\centering
\includegraphics[width=.8\textwidth]{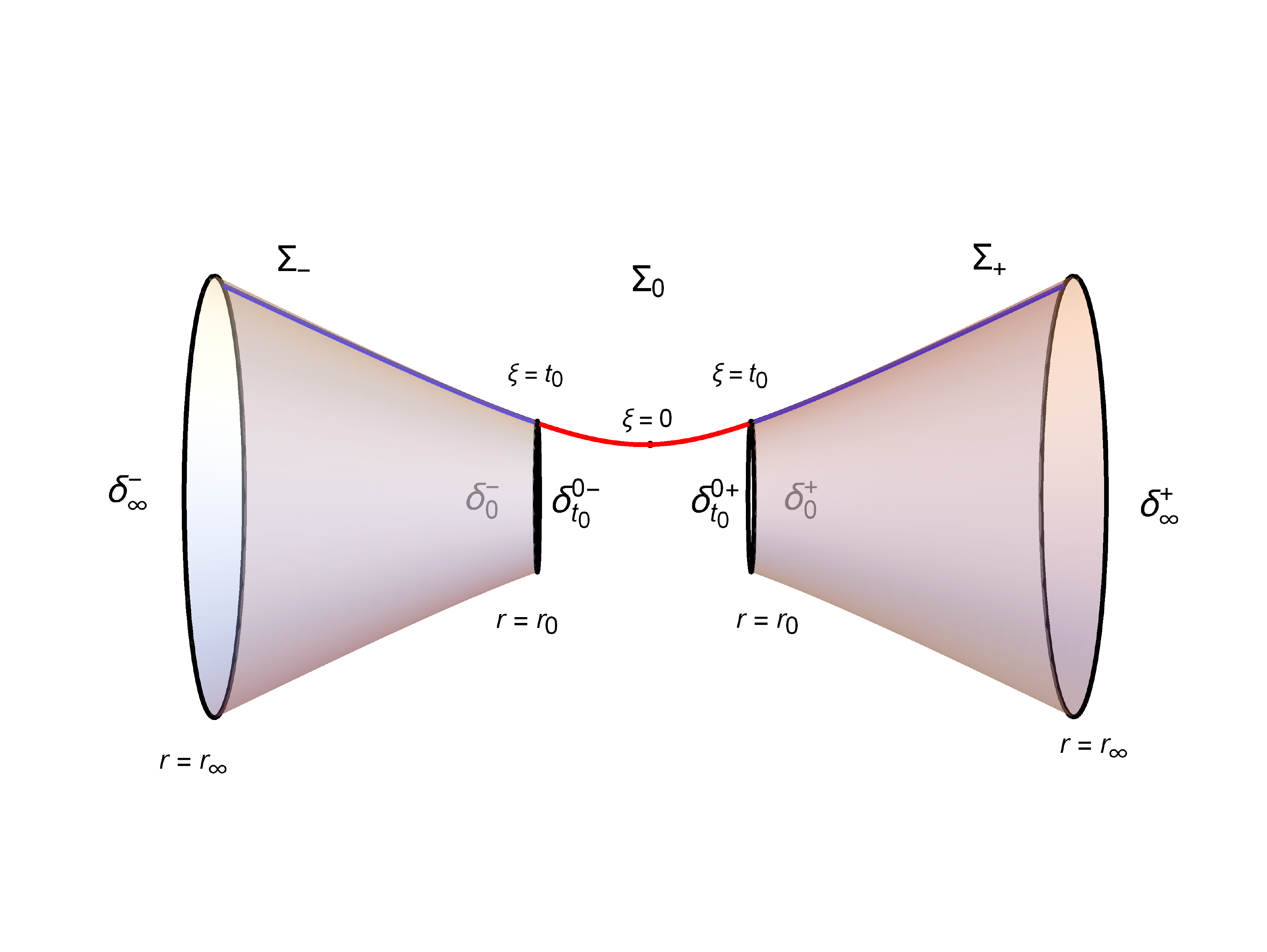}
\caption{\sl Two-dimensional representation for $t_0\neq 0$, of the 2d surfaces $\delta_{\rho}^{-}$ and $\delta_{\rho}^{+}$, and the $\Sigma_0$ slice emmbeded in four dimensions.}\label{extension3}
\end{figure}

The picture so far is: for $t_0=0$, see Fig. \ref{extension0}, while for $t_0\neq 0$, see Fig. \ref{extension3}. It is worth to emphasize that Fig. \ref{extension0} and Fig. \ref{extension3} are just 2d representations of two-dimensional surfaces embedded in $\mathbb{R}^4$.


Now we have the task of extending the portion of the $t_0$-slice, denoted by $\Sigma_0$ (the red line in Fig. \ref{Nice_Slice}). For this portion, the extension is less obvious since the variable time does not appear explicitly on $\Sigma_0$ as it does on $\Sigma_{-}$ and $\Sigma_{+}$. We shall call this surface $\delta_{\zeta}^0$, and its boundaries $\delta_{t_0}^{0-}$, and $\delta_{t_0}^{0+}$.

To extend $\Sigma_0$, we should note that the only time dependence of these slices appears at the boundaries. As mentioned, these slices only grow in time, but they are fixed at $\text{r}=\text{r}_0$.

First, let us parameterize the $\Sigma_0$ silces using a new parameter $\zeta$. On $\Sigma_0$ we could use the parametrization $\text{T}_1=\text{R}\ \text{cosh}({\frac{\zeta}{4\text{M}}})$, and $\text{X}_1=\text{R} \ \text{sinh}({\frac{\zeta}{4\text{M}}})$, with $-t_0\leq\zeta\leq t_0$. However, we find it more convenient to make the distinction $\Sigma_{0-}$ for $\text{X}_1\leq 0$, and $\Sigma_{0+}$ for $\text{X}_1\geq 0$, $\Sigma_{0}=\Sigma_{0-}\cup\Sigma_{0+}$; and use the following parametrization:
on $\Sigma_{0-}$
\bea\label{sigma0_equ_def1}
\text{T}_1 & = & +\text{R}\ \text{cosh}({\frac{\zeta}{4\text{M}}}),\nonumber\\
\text{X}_1 & = & -\text{R} \ \text{sinh}({\frac{\zeta}{4\text{M}}}) ,\quad 0\leq\zeta\leq t_0;
\eea
while on $\Sigma_{0+}$
\bea \label{sigma0_equ_def2}
\text{T}_1 & = & +\text{R}\ \text{cosh}({\frac{\zeta}{4\text{M}}}),\nonumber\\
\text{X}_1 & = & +\text{R} \ \text{sinh}({\frac{\zeta}{4\text{M}}}) ,\quad 0\leq\zeta\leq t_0\ ;
\eea
see Fig. \ref{extension3}. Note the dependence of $t_0$, on the boundaries $\delta_{t_0}^{0-}$, and $\delta_{t_0}^{0+}$, of $\Sigma_0$.

Plugging \eqref{sigma0_equ_def1} and \eqref{sigma0_equ_def2} in \eqref{L_section_metric} we can find the induced  metric on $\Sigma_{0}=\Sigma_{0-}\cup\Sigma_{0+}$. In the coordinates $(\zeta,\theta,\phi)$, it is given by
\be
\text{ds}^2=\frac{2\text{M}}{\text{r}_{0}}\text{e}^{-\frac{\text{r}_{0}}{2\text{M}}}\text{R}^2\text{d}\zeta^2+\text{r}_{0}^2\text{d}\Omega^2\label{metricsigma0}.
\ee
On the other hand, the induced  metric at $\text{r}=\text{r}_0$, i.e., on the boundaries $\delta_0^-$ and $\delta_0^+$, which would correspond to the boundaries  $\delta_{t_0}^{0-}$ and $\delta_{t_0}^{0+}$, respectively  (see Fig. \ref{extension3}) is
\be
\text{ds}^2=-\frac{2\text{M}}{\text{r}_{0}}\text{e}^{-\frac{\text{r}_{0}}{2\text{M}}}\text{R}^2\text{d}\tau^2+\text{r}_{0}^2\text{d}\Omega^2\label{metricatt0}.
\ee
Although $\text{r}=\text{r}_0$, is a coordinate singularity for the metric in the form \eqref{complexmetric}, \eqref{metricatt0} can be obtained directly from \eqref{complexmetric}, or more easily from the Wick rotated version of \eqref{metricrtcoor} and the relation \eqref{condi_r}.

With this in mind we conclude that the extension of $\Sigma_0$ will be driven only by the boundaries values of the metric on $\delta_{t_0}^{0-}$, and $\delta_{t_0}^{0+}$, which match the boundaries values of the metric on $\delta_{0}^{-}$ and $\delta_{0}^{+}$ \eqref{metricatt0}, respectively. Also, by the condition that at $\tau=0$, the induced metric of the complex extension matches the induced metric on the real slice $\Sigma_0$ \eqref{metricsigma0}, see Fig. \ref{extension3}. Therefore, the solutions we are seeking are those four-geometries that satisfy the boundary conditions \eqref{metricsigma0} and \eqref{metricatt0}.

The ansatz for the surface $\delta_{\zeta}^0$ takes the form:
for the extension of $\Sigma_{0-}$,
\bea
\text{T} & = & +\text{R}(\tau,\zeta)\ \text{cosh}({\frac{\zeta-\text{i}\tau}{4\text{M}}}),\nonumber\\
\text{X} & = & -\text{R}(\tau,\zeta) \ \text{sinh}({\frac{\zeta-\text{i}\tau}{4\text{M}}}) \ ,\quad 0\leq\zeta\leq t_0,\label{ ansatz1}
\eea
while for the extension of $\Sigma_{0+}$,
\bea
\text{T} & = & +\text{R}(\tau,\zeta)\ \text{cosh}({\frac{\zeta-\text{i}\tau}{4\text{M}}})\nonumber\\
\text{X} & = & +\text{R}(\tau,\zeta) \ \text{sinh}({\frac{\zeta-\text{i}\tau}{4\text{M}}}) \ , \quad 0\leq\zeta\leq t_0\ ,\label{ ansatz2}
\eea
where $\text{R}(\tau,\zeta)$, is a real function and $\tau\sim\tau+8\pi\text{M}$.

The vacuum solution of the Einsten's equations on $\delta_{\zeta}^0$ has the same form as in \eqref{metriccomplex}, but $(\text{T},\text{X})$ are defined on the complex surface $\delta_{\zeta}^0$ given by \eqref{ ansatz1} and \eqref{ ansatz2}. Plugging \eqref{ ansatz1} and \eqref{ ansatz2} in \eqref{metriccomplex} we get a family of complex metrics, where
\be
\text{r}=2\text{M}\Big(1+\text{W}_0\big(-\frac{\text{R}(\tau,\zeta)^2}{\text{e}}\big) \Big)\in\mathbb{R}.\label{r_on_delta0_zeta}
\ee
Naively one may think that $\text{R}(\tau,\zeta)=\text{R}$, is the simplest solution. The obstacle to such a choice is that a constant $\text{R}(\tau,\zeta)$, leads to a non invertible metric.

In order to avoid possible metric singularities \footnote{Notice that if $\text{R}(\tau,\zeta)=1$, \eqref{r_on_delta0_zeta} would vanish. Recal that $\text{W}_0(-\text{e}^{-1})=-1$, and $\text{r}=0$, is a singular point for the metric \eqref{metriccomplex}.} on $\delta_{\zeta}^0$ and, as we necesarily need a non constant function $\text{R}(\tau,\zeta)$, now we have to move the conditions on $\text{R}$ to the function $\text{R}(\tau,\zeta)$. In other words, we consider only solutions with small (small enough but not infinitesimal) deviations from the constant value $\text{R}$, i.e., $\text{R}(\tau,\zeta)=\text{R}+s(\tau,\zeta)<<1$, with $s(\tau,\zeta)\sim 0$.

Continuity and consistency with \eqref{metricsigma0} and \eqref{metricatt0} requires
\begin{enumerate}
\item $\text{R}(\tau,\zeta)=\text{R}(\tau+8\pi\text{M},\zeta)\in \mathbb{R}$,
\item $\text{R}(0,\zeta)=\text{R}$,
\item $\text{R}(\tau,t_0)=\text{R}$,
\item $\partial_{\zeta}\text{R}(0,\zeta)=0$,
\item $\partial_{\tau}\text{R}(\tau,t_0)=0$.
\end{enumerate}

\begin{figure}[h]
\centering
\includegraphics[width=.8\textwidth]{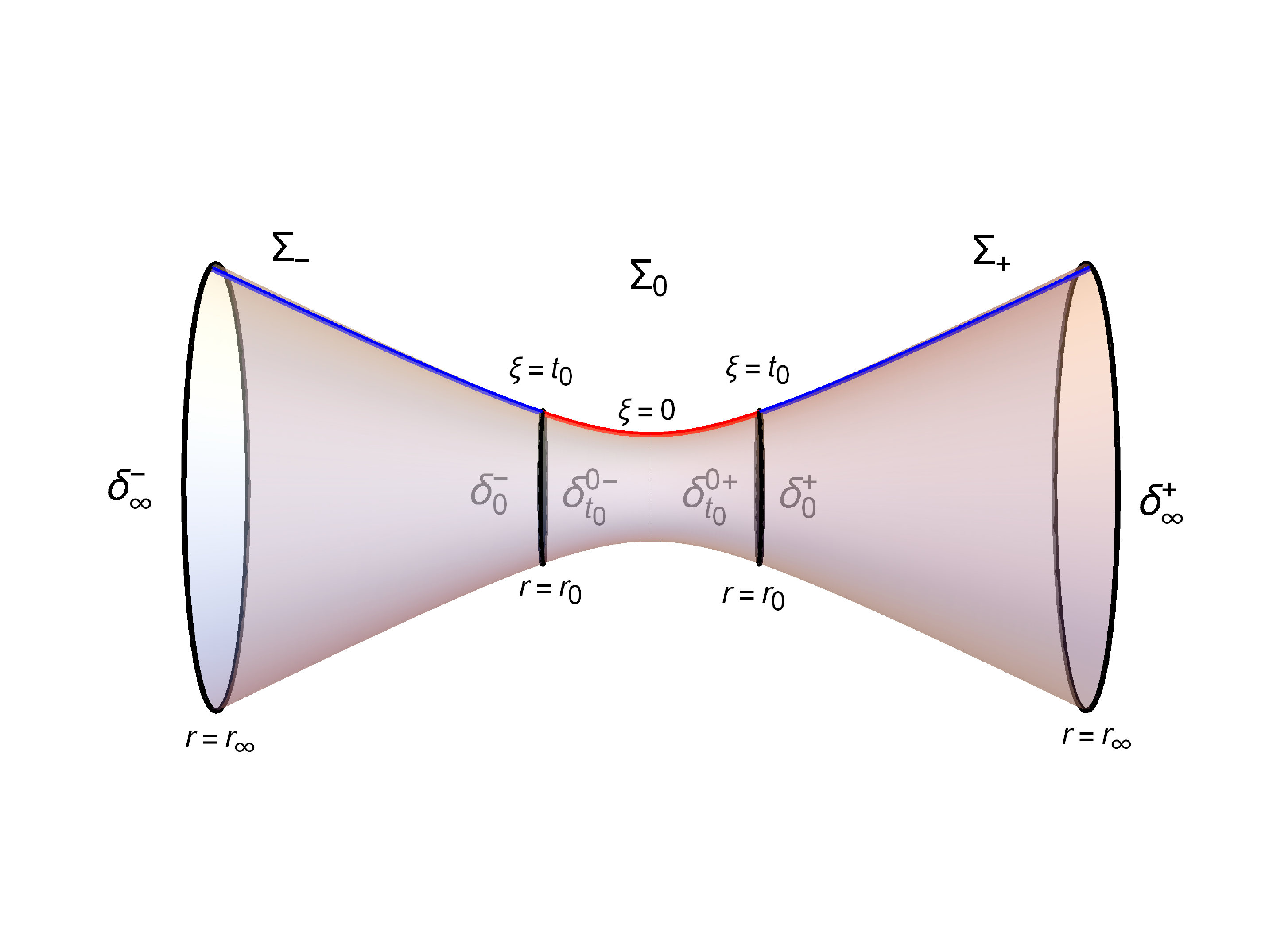}
\caption{\sl Full picture of the manifold obtained after the complex extension of the $t_0$-slice.}\label{extension2}
\end{figure}

Now we are in a condition to represent the full picture of the geometry of the complex extension of a nice slice Fig. \ref{extension2}.

\subsection{Extrinsic curvature}\label{subsec:3.1}

We have found a family of manifolds that matches continuously with $\delta_{\rho}^{-}$ and $\delta_{\rho}^{+}$, but this is not the end of the story. In order to fully determine the solution we also have to impose smoothness at the matching surfaces. For that we must compute the extrinsic curvature defined as
\be
\text{K}_{ab}=-(\partial_a\gamma_b^{\mu}\hat{\text{n}}_{\mu}+\Gamma^{\mu}{}_{\nu\rho} \hat{\text{n}}_{\mu} \gamma_a^{\nu}\gamma_b^{\rho}).
\ee
As we are interested in computing the extrinsic curvature at a constant value of a coordinate, either $\text{r}=\text{r}_0$, or $\zeta=t_0$, or on the asymptotic boundaries at $\text{r}=\text{r}_{\infty}$, $\text{K}_{ab}$, reduces to
\be
\text{K}_{ab}=-\Gamma^{\mu}{}_{\nu\rho} \hat{\text{n}}_{\mu} \gamma_a^{\nu}\gamma_b^{\rho},
\ee
where
\bea
\gamma_{1}^{\mu} & = & (1,0,0,0),\nonumber \\
\gamma_{2}^{\mu} & = & (0,0,1,0),\nonumber \\
\gamma_{3}^{\mu} & = & (0,0,0,1),\nonumber \\
\hat{\text{n}}_{\mu} & = & \frac{(0,1,0,0)}{\sqrt{\epsilon \ g^{22}}},
\eea
with $\epsilon=\pm1$, according to the signature of the metric.

In what follows we use the superscripts $0-$, $0+$, $-$ and $+$, in the tensor $\text{K}_{ab}$ to indicate on which boundary we are computing the extrinsic curvature according to  the superscripts of $\delta_{t_0}^{0-}$, and $\delta_{t_0}^{0+}$, $\delta_{0}^{-}$, and $\delta_{0}^{+}$ respectively.

First, for consistency, we have checked that at $\zeta=0$ both spaces \eqref{ ansatz1}, and \eqref{ ansatz2} match smoothly, see Fig. \ref{extension2}
\be
\text{K}_{ab}{}_{{\big |}_{\zeta=0}}=\text{K}_{ab}{}_{{\big |}_{\zeta=0}}.
\ee
The extrinsic curvature on the boundaries $\delta_{t_0}^{0-}$, and $\delta_{t_0}^{0+}$, is given by
\be
\text{K}_{ab}^{0-}{}_{{\big |}_{\zeta=t_0}}=\text{K}_{ab}^{0+}{}_{{\big |}_{\zeta=t_0}} \nonumber
\ee
\begin{align}\nonumber
= \text{sign}(\partial_{\zeta}\text{R}(\tau,t_0))(2\text{M}-\text{r}_0)^{\frac{1}{2}} \times \\
\text{diag} \Big(\frac{\text{M}}{\text{r}_0^{\frac{5}{2}}}\big(1-\frac{3\text{r}_0^2}{\text{R}}\partial_{\tau}\partial_{\tau}\text{R}(\tau,t_0)\big)\ , \ \text{r}_0^{\frac{1}{2}}\ , \ \text{r}_0^{\frac{1}{2}}\text{sin}^2(\theta) \Big).
\end{align}
While on the boundaries $\delta_{0}^{-}$ and $\delta_{0}^{+}$, is
\be
\text{K}_{ab}^{-}{}_{{\big |}_{\text{r}=\text{r}_0}}=\text{K}_{ab}^{+}{}_{{\big |}_{\text{r}=\text{r}_0}} = -(2\text{M}-\text{r}_0)^{\frac{1}{2}} \text{diag}
\Big(\frac{\text{M}}{\text{r}_0^{\frac{5}{2}}}\ , \ \text{r}_0^{\frac{1}{2}}\ , \ \text{r}_0^{\frac{1}{2}}\text{sin}^2(\theta) \Big).
\ee

Requering that on the boundaries $\delta_{t_0}^{0-}$, $\delta_0^{-}$, and $\delta_{t_0}^{0+}$, $\delta_0^{+}$, see Fig. \ref{extension2}, both spaces match smoothly
\bea
\text{K}_{ab}^{0-}{}_{{\big |}_{\zeta=t_0}} & = & \text{K}_{ab}^{-}{}_{{\big |}_{\text{r}=\text{r}_0}},\nonumber \\
\text{K}_{ab}^{0+}{}_{{\big |}_{\zeta=t_0}} & = & \text{K}_{ab}^{+}{}_{{\big |}_{\text{r}=\text{r}_0}},
\eea
leads to the extra conditions
\begin{enumerate}
\setcounter{enumi}{5}
\item $\partial_{\tau}\partial_{\tau}\text{R}(\tau,t_0)=0$,
\item $\partial_{\zeta}\text{R}(\tau,t_0)\leq0$.
\end{enumerate}

The functions that satisfy the conditions listed above are
\be
\text{R}(\tau,\zeta)=\text{R}+\sum_{n=1}^{\infty}a_{n}(\zeta)\text{sin}(\frac{n}{4\text{M}}\tau)\ ,\quad a_{n}(t_0)=0;\label{R}
\ee
with $a_n(\zeta)$, such that $\partial_{\zeta}\text{R}(\tau,t_0)\leq0$, holds, and $\text{R}(\tau,\zeta)<<1$.

Despite the metric \eqref{metriccomplex} or \eqref{complexmetric} on the extension \eqref{2dsurface} is complex, the Gibbons–Hawking–York term is
\be
\int\limits_{\text{r}=\text{r}_{\infty}} \text{K}\sqrt{h}\text{d}x^3=-32\pi^2\text{i}\text{M}(2\text{r}_{\infty}-3\text{M}).
\ee
Therefore, the only contribution to the action \eqref{EHaction} is
\be
\text{I}[g_c](\beta)=2\times (8\pi)^{-1}\int\limits_{\text{r}=\text{r}_{\infty}\rightarrow\infty} [\text{K}]\sqrt{h}\text{d}x^3=2\times 4\pi \text{i}\text{M}^2= \text{i}  \ \frac{\beta^2}{8\pi},\label{2part_func}
\ee
where the factor $2$  appears because there are two asymptotics boundaries, $\delta_{\infty}^{-}$ and $\delta_{\infty}^{+}$. Like in the Hawking's calculation in section \ref{sec:Hawking's calculation}, using thermodynamics arguments \eqref{2part_func} leads to
\be
\text{S}_{BH}=\frac{A}{2}.\label{expected_entropy2BH}
\ee

We have computed the thermodynamics entropy of a black hole on a nice slice; however, we do not know yet which state leads to such an entropy. Before moving to the next section, where we discuss the density matrix interpretation of the calculation presented above, we shall point out another feature of the geometry we have obtained.

\begin{figure}[h]
\centering
\includegraphics[width=.6\textwidth]{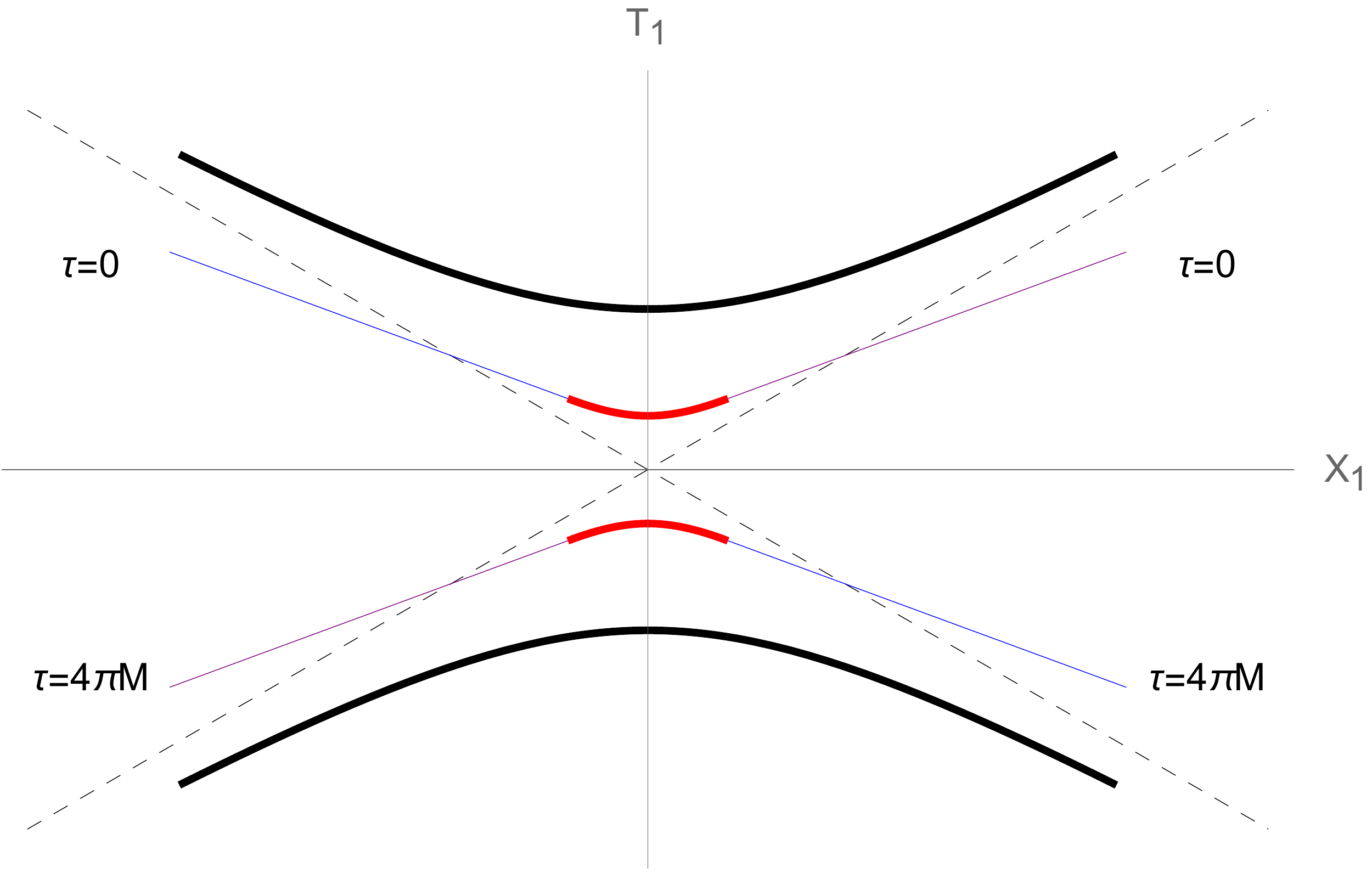}
\caption{\sl Intersections of the complex-extended manifold with the real space. The blue lines represent the intersections of the $\delta_{\rho}^{-}$ surface with the real space. At the same time, the magenta lines represent the intersections of $\delta_{\rho}^{+}$. For instance, starting from the blue line on the upper left at $\tau=0$, and evolving it in complex time up to $\tau=4\pi\text{M}$, it reaches the blue line on the lower right. Similarly, for the red lines.}\label{Analisis}
\end{figure}

From \eqref{2dsurface} and \eqref{R} we can see this geometry intersects the Lorentzian space in two differents surfaces. The surface $\tau=0$, which corresponds to the slice $t=t_0$, on the real space, and the surface $\tau=4\pi\text{M}$, which corresponds to the $\text{T}_1$-reflected slice of $t=t_0$, see Fig. \ref{Analisis}.

Let us stress one more point. In this section, we have considered that the $\Sigma_0$ slice extension leads to a manifold that is topologically equivalent to two cylinders joined at their boundaries, with opposite orientation (similar to Fig. \ref{extension0} but elongated in the $\zeta$ direction). In principle, we could consider contributions from manifolds with higher genus topologies. As long as the $\Sigma_0$ slice belongs to these manifolds, the definition of state on the Lorentzian space will remain untouched. Without considering the matter fields, there will not be a semiclassical contribution to the action coming from these manifolds because they would be solutions of the vacuum Einstein's equations. However, the situation would be different if matter fields are taken into account.

\section{Density matrix interpretation}\label{sec:5}

At this point one might be tempted to define a density matrix $\rho\big[ h^{+}_{ij},\phi^{+}_0 ; h_{ij}^{-},\phi_0^{-} \big]=\Psi\big[ h^{+}_{ij},\phi^{+}_0 \big]\Psi^{*}\big[h_{ij}^{-},\phi_0^{-} \big]$,  and associate it to the geometry above to describe a semiclassical  state.  $\Psi^{*}\big[h_{ij}^{-},\phi_0^{-} \big]$ and $\Psi\big[ h^{+}_{ij},\phi^{+}_0 \big]$ would be defined on two disjoint geometries with boundary values on the surfaces $\tau=0^{-}$, $(h_{ij}^{-},\phi_0^{-})$; and $\tau=8\pi\text{M}\sim 0^{+}$, $(h_{ij}^{+},\phi_0^{+})$. 

\begin{figure}[h]
\centering
\includegraphics[width=.8\textwidth]{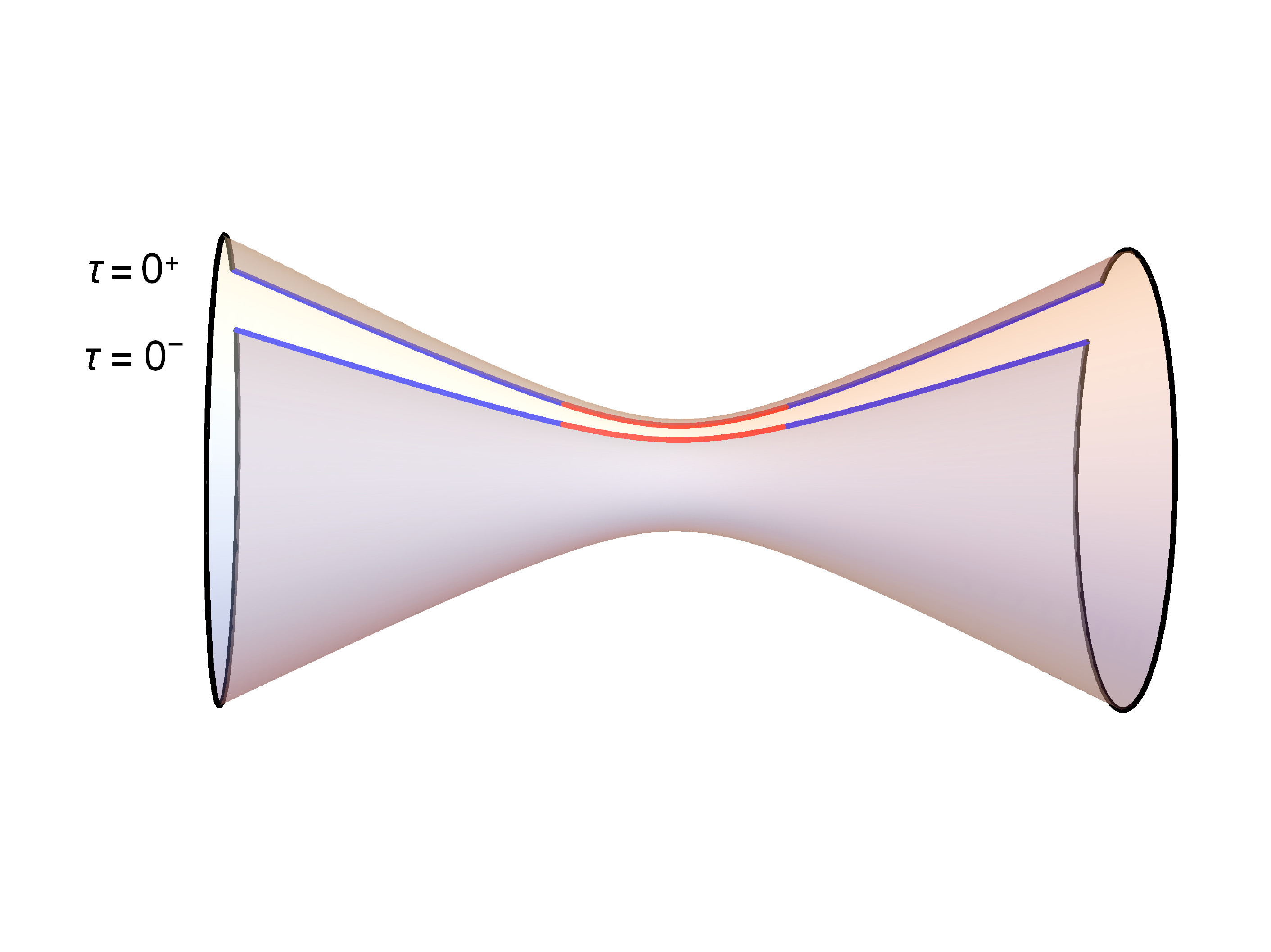}
\caption{\sl Representation of the complex-extended geometry that connects the boundary $\tau=0^{-}$, with the boundary $\tau=0^{+}$.}\label{PF_cut}
\end{figure}

The issue is that for this geometry the  associated  density matrix  can not be factorized. To see this, we can just evolve the slice $t=t_0$, in imaginary time $\tau$, and note that the slices $\tau=0^{-}$, and $\tau=8\pi\text{M}\sim 0^{+}$, Fig. \ref{PF_cut} are connected by a surface.  In other words, this geometry contains noncontractible loops crossing the slice $t=t_0$  \cite{Page:1986vw}. 

The reader can contrast with Fig. \ref{E_state2} where the state is pure, and the manifold associated to the density matrix is composed of two disjoint semi-disk. Note that in Fig. \ref{E_state2},  in contrast with  Fig. \ref{PF_cut}, the boundaries of $\rho$ are dissconected. When the entries (or the boundaries) of the density matrix are connected, it describes a mixed state, see for instance \cite{Page:1986vw} for a  disscusion about this.

Moreover, this geometry intersects the  Lorentzian  space in two differents surfaces Fig. \ref{Analisis}. In other words, this geometry divides the real space in more than two parts. So, as discused in \cite{Hawking:1986vj, Page:1986vw},  the semiclassical state described by the geometry above corresponds to a global  mixed state with an associated density matrix of the form
\be
\rho\big[h^{+}_{ij},\phi^{+}_0 ; h_{ij}^{-},\phi_0^{-} \big]=\sum_{m,n}C_{mn}\Psi_m\big[ h^{+}_{ij},\phi^{+}_0 \big] \Psi^{*}_n\big[h_{ij}^{-},\phi_0^{-} \big],\label{den_mix}
\ee
fulfilling the equation \eqref{WDW2}, where $C_{mn}$ does not factorize, i.e., $C_{mn}\neq c_mc_n$.
We would like to stress that \eqref{den_mix} is not the density matrix associated to the thermofield double (TFD) of the HH state \cite{Maldacena:2001kr}. The wave functionals $\Psi_m\big[ h^{+}_{ij},\phi^{+}_0 \big]$, and $\Psi^{*}_n\big[h_{ij}^{-},\phi_0^{-} \big]$, in \eqref{den_mix}  are defined on the whole nice slice $t=t_0$, Fig. \ref{Nice_Slice}, and not only on half of the space as in the TFD.  We refer to section $3.3$ and $4.8$ of \cite{Harlow:2014yka} and references therein for a discussion about the TFD in the context of BH's. 

The state associated to \eqref{den_mix} is not pure, but yet after tracing over the boundary values on the $t_0$-slice we get the expected entropy, as shown in the previous section in equation \eqref{2part_func}, and disscussed in section \ref{sec:2} in equation \eqref{Tr_mixed_den},  i.e., 
\begin{align}\label{Zconne_g_c}
\text{Z}(\beta)=\text{Tr}\big[\rho \big]=\int{D}h_{ij}\ \sum_{m,n}C_{mn}\Psi_m\big[ h_{ij} \big] \Psi_n\big[h_{ij}\big]=\ \ \ \ \ \ \ \ \ \ \ \ \ \ \ \ \ \ \ \ \nonumber \\
\sum_{\begin{matrix}\text{disconnected}\\ +\ \text{connected}\end{matrix}}\int \text{D}g\text{D}\phi \text{exp}\big(\text{i}\text{I}[g,\phi]\big)\sim \text{exp}\big(\text{i}\text{I}[g_{c}]\big)\Big{|}_{\begin{matrix}\text{connected}\\ \text{only} \end{matrix}}=\text{exp}\big[- \frac{\beta^2}{16\pi}\big],
\end{align}
where to match the calculation in the previous section we have removed the matter fields appearing in \eqref{den_mix}.  We would like to stress that $g_c$  in \eqref{Zconne_g_c}  is the metric we have built in the previous section.

\begin{figure}[h]
\centering
\includegraphics[width=.8\textwidth]{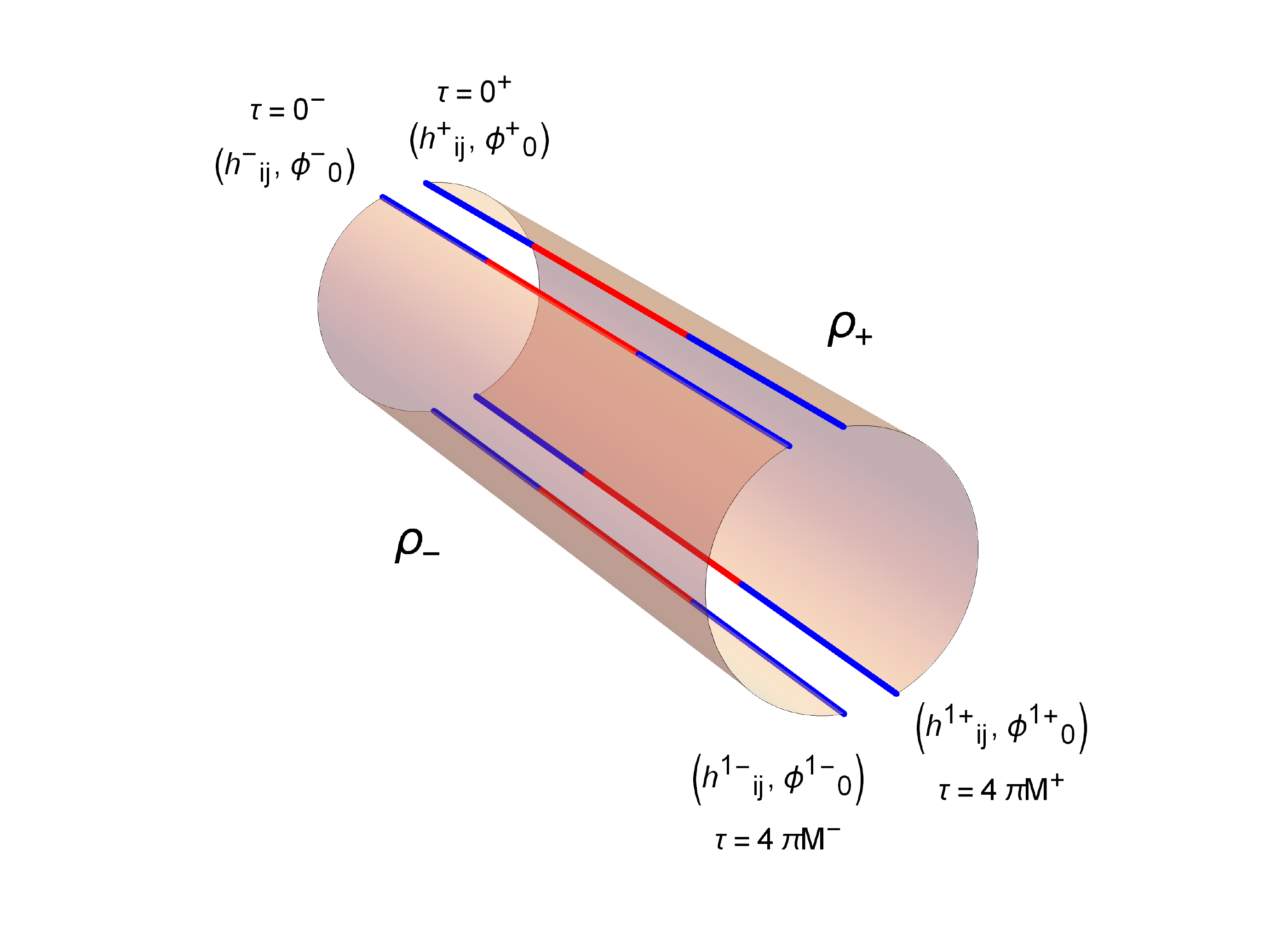}
\caption{\sl Representation of the manifolds associated to $\rho^{-}$ and $\rho^{+}$.}\label{dens_fact1}
\end{figure}

Interestingly enough, \eqref{den_mix} factorizes in two density matrices \cite{Hawking:1986vj}
\be
\rho\big[h_{ij}^{+},\phi_0^{+}; h^{-}_{ij},\phi^{-}_0 \big]=
\int\text{D}h_{ij}^{1}\text{D}\phi_0^{1}\rho_{+}\big[h_{ij}^{+},\phi_0^{+}; h^{1}_{ij},\phi^{1}_0 \big]\rho_{-}\big[h_{ij}^{1},\phi_0^{1}; h^{-}_{ij},\phi^{-}_0 \big].\label{den_fact}
\ee
The boundary values $(h_{ij}^1,\phi_0^1)$, match the value of the fields on the $\text{T}_1$-reflected slice of $t=t_0$, at $\tau=4\pi\text{M}$, as discussed in Fig. \ref{Analisis}. In this case we can see that $\rho_{-}$, and $\rho_{+}$, do not correspond to pure states since each one comes from a {\it connected} geometry Fig. \ref{dens_fact1}; and the trace over the non-observable boundary $\tau=4\pi\text{M}$, leads to \eqref{den_fact}.

\begin{figure}[h]
\centering
\includegraphics[width=.6\textwidth]{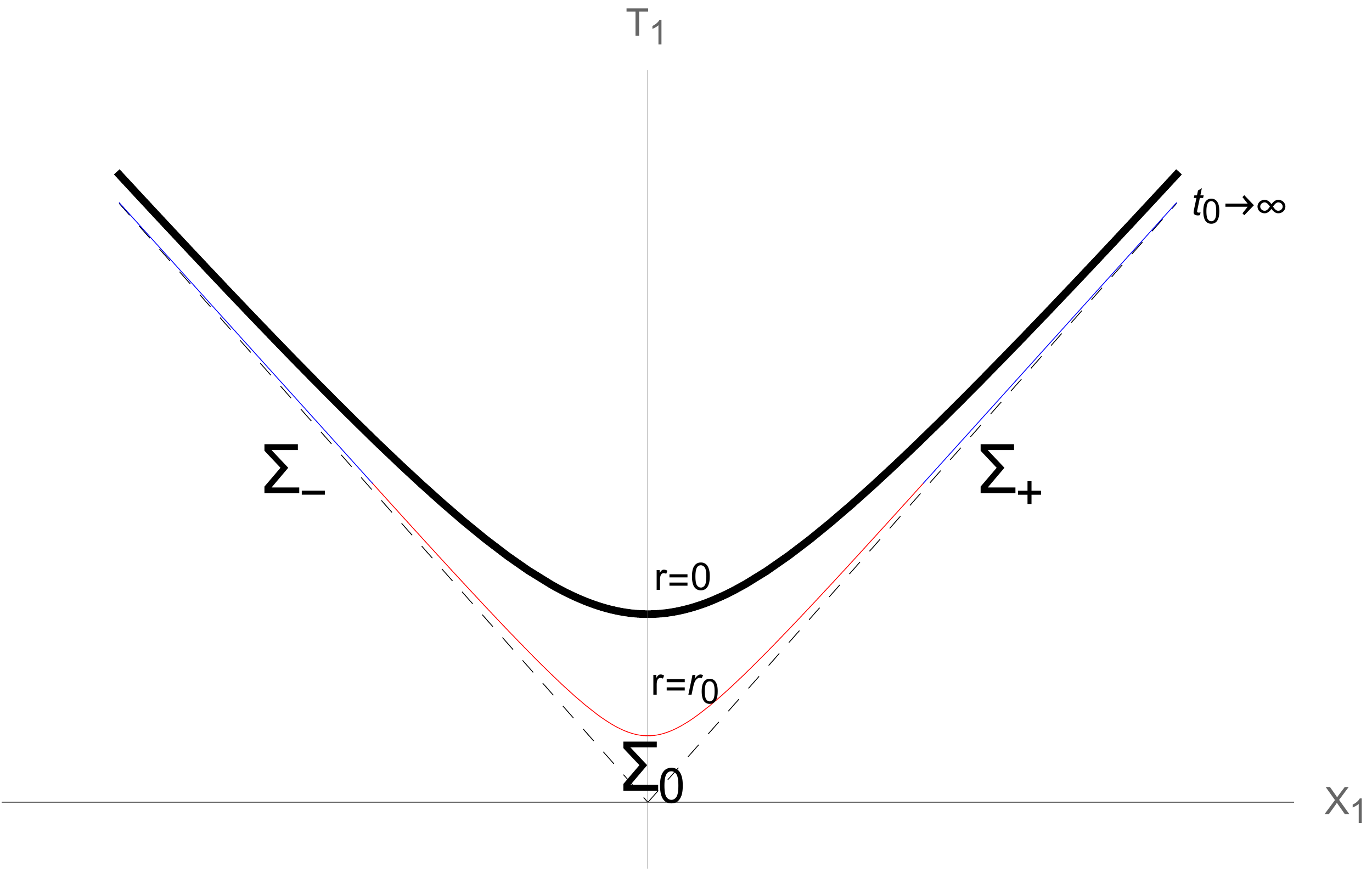}
\caption{\sl Schematic representation of the slice $t_0\rightarrow\infty$. Notice that when $t_0\rightarrow\infty$, the red line inside the horizon becomes infinitely long and $\Sigma_{-}$ and,
$\Sigma_{+}$ lie on the horizons, at null infinity.}\label{NI_nice_slice}
\end{figure}

Now we can regard $\rho_{-}$, and $\rho_{+}$, as transition amplitudes. For instance, $\rho_{+}$ could be seen as the transition amplitude from the state on the slice $t=-t_0$, ($\tau=4\pi\text{M}$) with values $(h_{ij}^{1-},\phi_0^{1-})$, to the state on the slice $t=t_0$, ($\tau=0$) with values $(h_{ij}^{-},\phi_0^{-})$. In fact, $\rho_{+}$ could be regarded as an $S$ matrix when $t_0\rightarrow\infty$. In the limit $t_0\rightarrow\infty$, the futute and past segments $\Sigma_{-}$ and $\Sigma_{+}$,  lie completely on null infinity,  $\mathcal{I}^{+}$ and  $\mathcal{I}^{-}$, see for instance  Fig. \ref{NI_nice_slice}. It is worth to stress that $\rho_{-}$, and $\rho_{+}$, satisfy the equations
\bea
\hat{\text{H}}\rho_{-}[ h^{1}_{ij},\phi_0^{1}; h^{-}_{ij},\phi_0^{-}] & = & \delta\big[h_{ij}^{1},h_{ij}^{-}\big]\delta\big[\phi_0^{1},\phi_0^{-}\big], \nonumber \\ 
\hat{\text{H}}\rho_{+}[ h^{+}_{ij},\phi_0^{+}; h^{1}_{ij},\phi_0^{1}] & = & \delta\big[h_{ij}^{+},h_{ij}^{1}\big]\delta\big[\phi_0^{+},\phi_0^{1}\big]\nonumber.
\eea

\section{Comments on the entanglement entropy and replica wormholes on a nice slice}\label{sec:6}

In this section, we shall point out the relation of our work, when extended to compute the entanglement entropy, with some recent proposals \cite{Almheiri:2019qdq, Penington:2019kki,Anegawa:2020ezn, Gautason:2020tmk, Hashimoto:2020cas, Hartman:2020swn}. Here we would see how following a slightly different logic, we arrive at the concept of replica wormhole. Although we do not consider the matter contribution in the following discussion, we give a prescription for how the entanglement entropy in QG should be computed for a four-dimensional  black hole on a nice slice.

For the state defined above we can compute its associated entanglement entropy. We will exemplify this calculation by posing the problem of computing the entaglement entropy for the segmets $\Sigma_{-}$ and $\Sigma_{+}$ on the silce $t_0<\infty$, see Fig. \ref{Nice_Slice}. Note that, at least, mathematically we can pose the problem on these segments for $t_0<\infty$. For them we have $\text{r}_0\leq\text{r}\leq \infty$, with $\text{r}_0<2\text{M}$. The subsequent discussion also applies to the more physical scenery where the segments are $\text{r}_1\leq\text{r}\leq \infty$, with $\text{r}_1>2\text{M}$. It also applies for $t_0\rightarrow\infty$, in which case  the segments $\Sigma_{-}$ and $\Sigma_{+}$ sit completely on null infinity.

\begin{figure}[h]
\centering
\includegraphics[width=.7\textwidth]{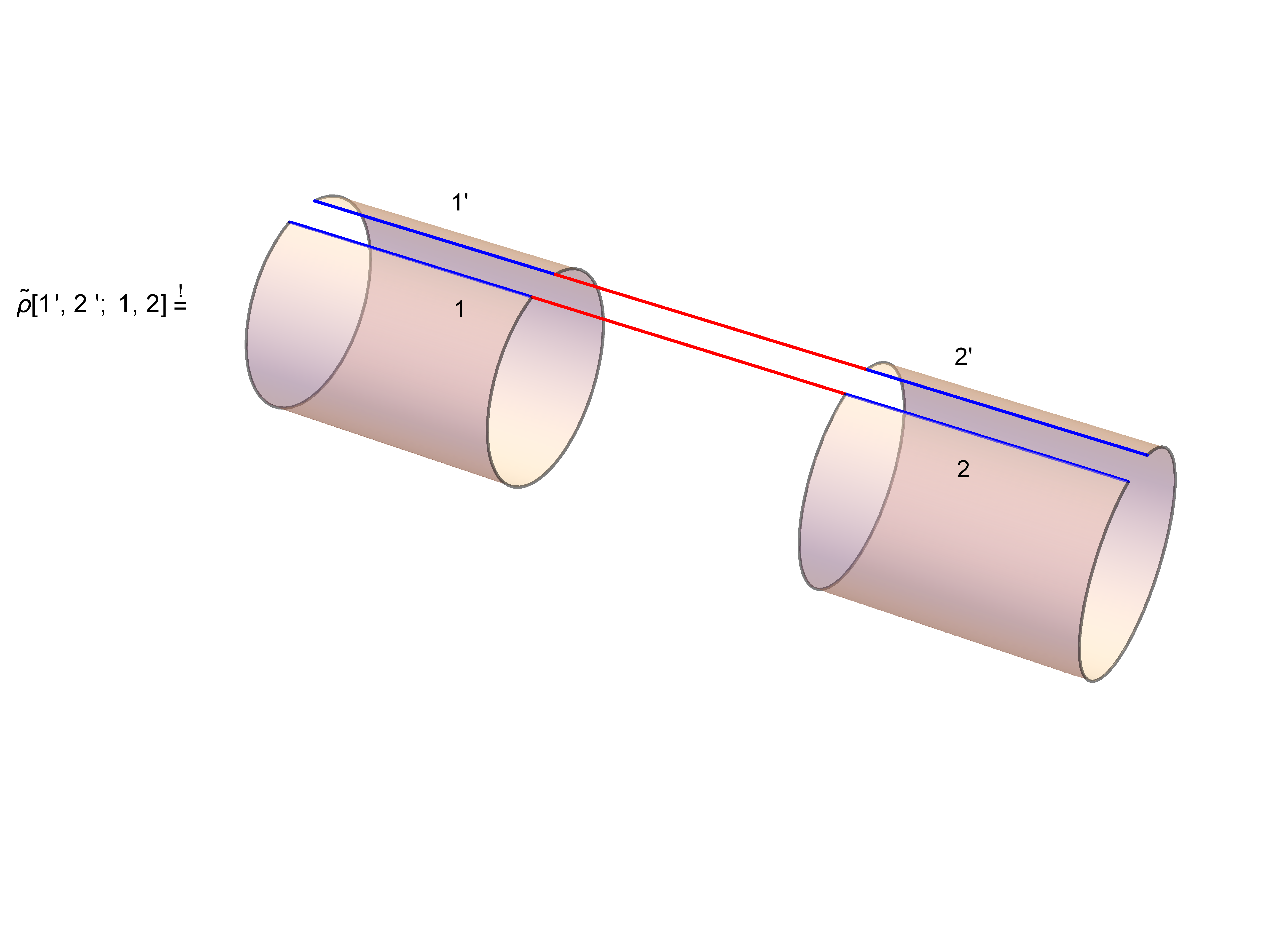}
\caption{\sl Complex extension of the segments $\Sigma_{-}$ and $\Sigma_{+}$, and geometric representation of the reduced density matrix. Here the reduced density matrix has not been fully specified yet. To fully specify it, we must fill in the geometry in between the two cylinders.}\label{dif_density_to_fill_in}
\end{figure}

To address this calculation, we must first define the replica manifold of this geometry. We can start by defining the reduced density matrix $\tilde{\rho}[1',2'; 1, 2]$ associated to $\Sigma_{-}$ and $\Sigma_{+}$.
To build this object first, we perform the complex extension on the segments $\Sigma_{-}$ and $\Sigma_{+}$, Fig. \ref{dif_density_to_fill_in}.
Then we should fill in the geometry for the extension of the $\Sigma_0$ slice. The symbol $\stackrel{!}{=}$, in the definition of the density matrix in Fig. \ref{dif_density_to_fill_in} indicates that $\tilde{\rho}[1',2'; 1, 2]$, has not been fully specified yet. Recall $t$ does not appear explicitly on $\Sigma_0$, and this slice does not evolve forward in time, it only grows. The extension of it is determined only by the metric's boundaries values on $\delta^{-}_0$ and $\delta^{+}_0$; and the induced metric on $\Sigma_0$. To fully specify the reduced density matrix, we have to fill in the geometry in between the two cylinders in Fig. \ref{dif_density_to_fill_in}, as we did in the previous section.

\begin{figure}[h]
\centering
\includegraphics[width=.7\textwidth]{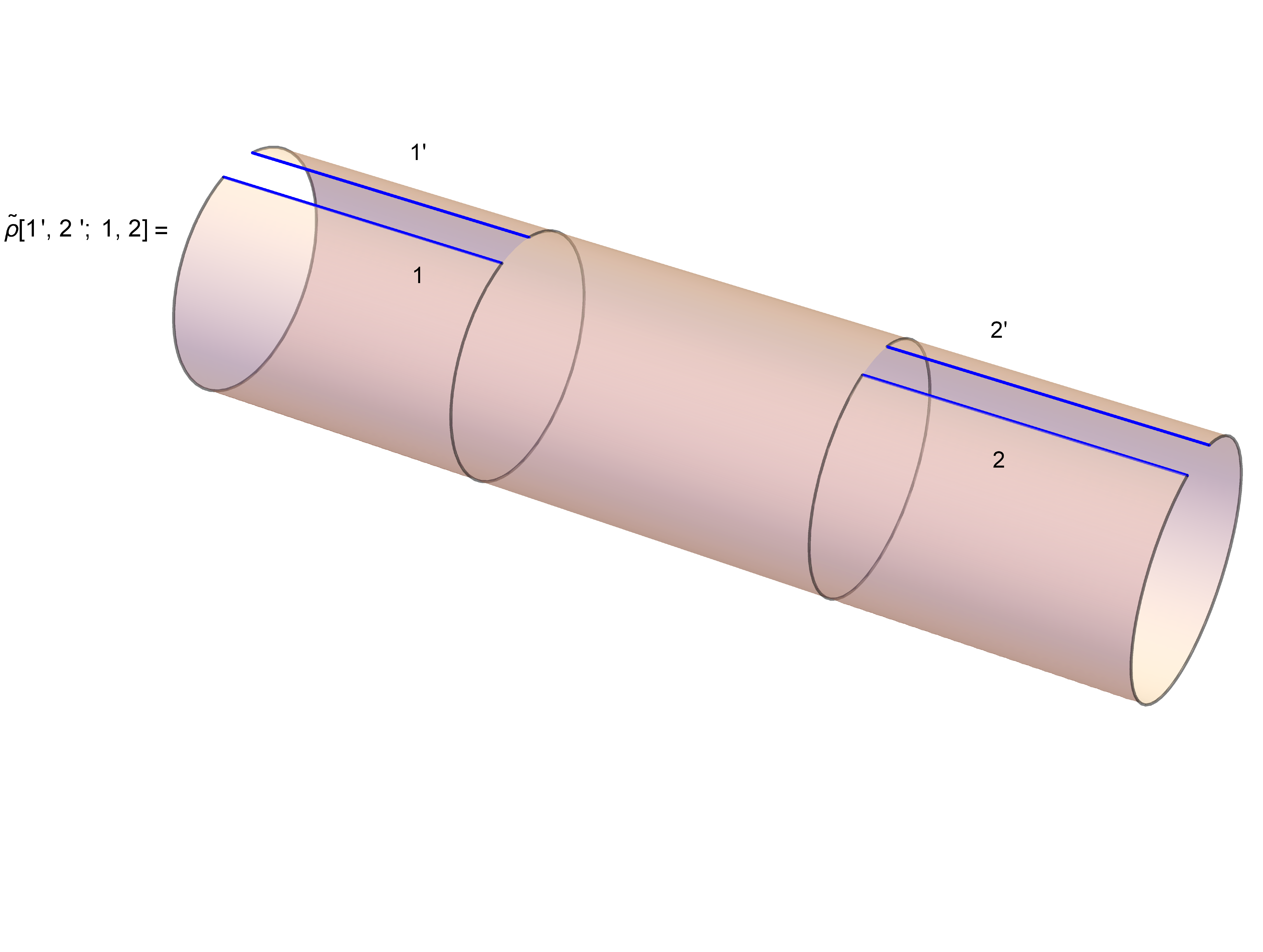}
\caption{\sl Geometric representation of the reduced density matrix.}\label{dif_density}
\end{figure}

The geometric representation of the reduced density matrix is depicted in Fig. \ref{dif_density}.
This reduced density matrix can be obtained by taking the partial trace of the density matrix defined in the previuos section over the degrees of freedom on $\Sigma_0$ (red slice, see for instance Fig. \ref{PF_cut}), i.e., $\tilde{\rho}=\text{Tr}_{\Sigma_0}[\rho]$. In this way the partition function would be $\text{Z}=\text{Tr}_{\Sigma_{-}\cup\Sigma_{+}}[\tilde{\rho}]$.

Using $\tilde{\rho}$, we can compute the density matrix of the replicated manifold. However, this construction comes with a caveat, and extra care is needed when we apply it to construct and associate $\tilde{\rho}^{n}$ to the replicated manifold. We should remember that there is an ambiguity when extending the $\Sigma_0$ slice. To see the consequences of such ambiguity, let us construct the manifold associated to $\tilde{\rho}^2$.

\begin{figure}[h]
\centering
\includegraphics[width=.8\textwidth]{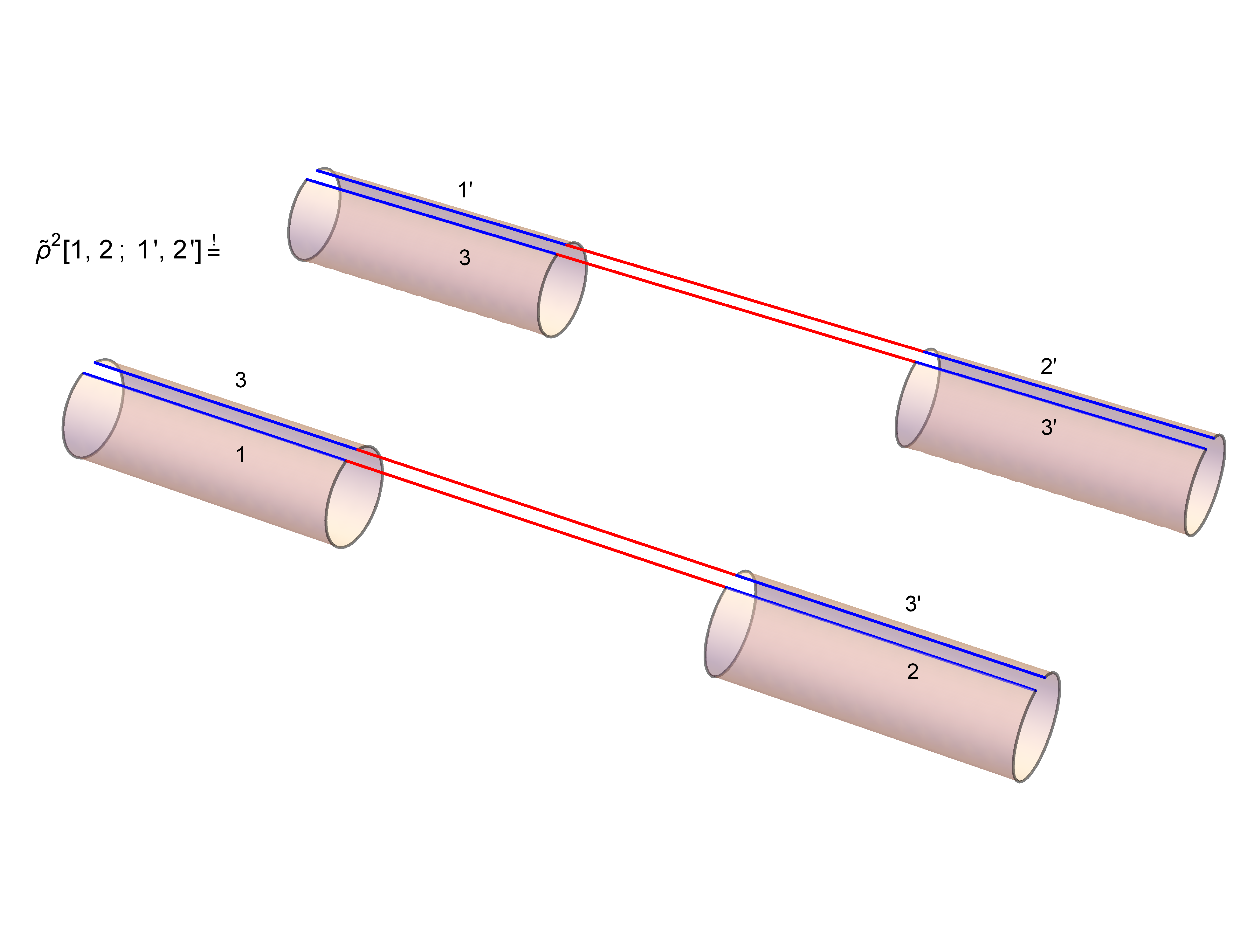}
\caption{\sl Complex extension of two copies of the segments $\Sigma_{-}$ and $\Sigma_{+}$, and geometric representation of the reduced density matrix for the replicated manifold. Here the reduced density matrix associated with the replicated manifold has not been fully specified yet. To fully specify it, we must fill in the geometry in between the four cylinders.}\label{rho2_to_fill_in}
\end{figure}

\begin{figure}[h]
\centering
\includegraphics[width=.8\textwidth]{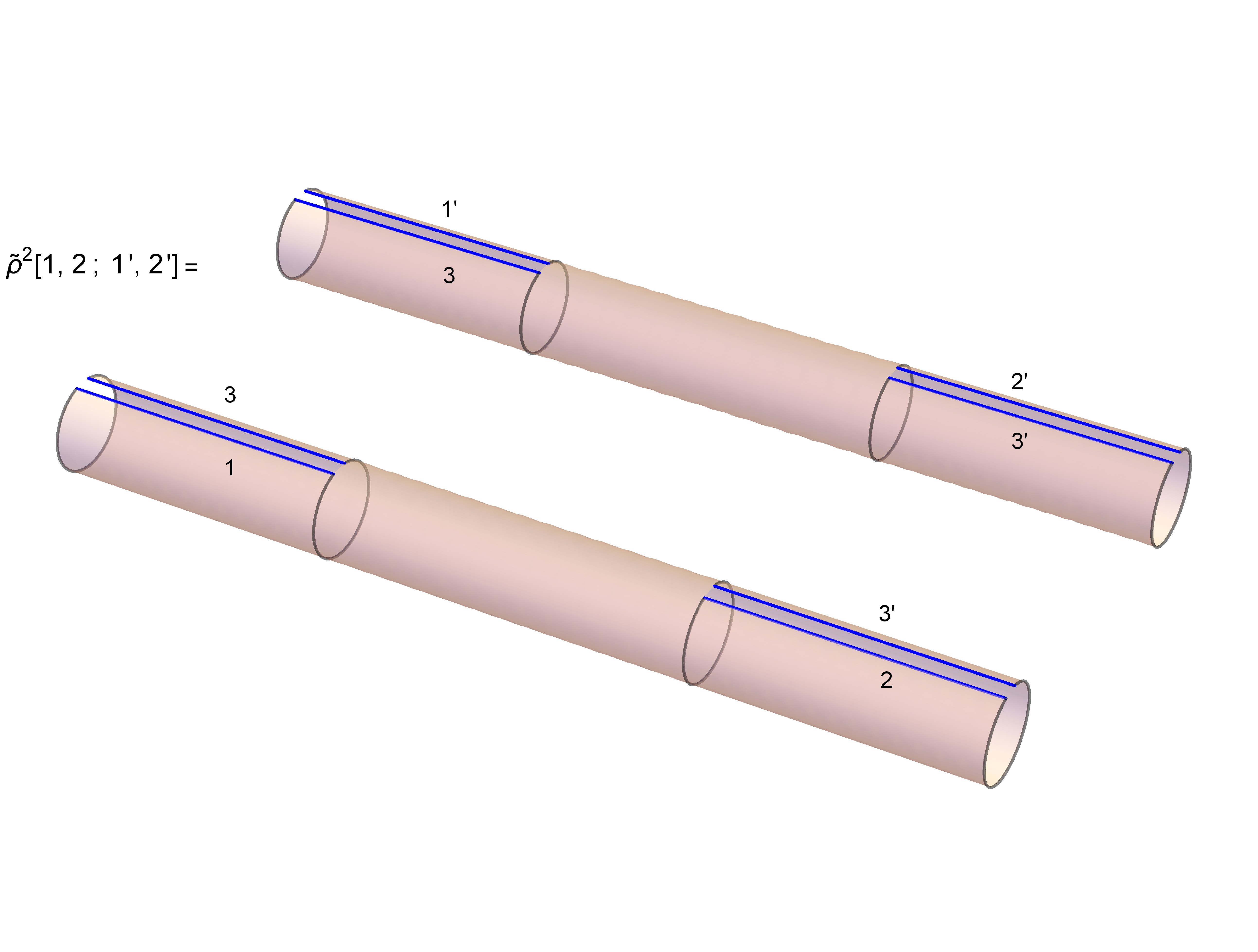}
\caption{\sl Geometric representation of the {\it genuine} $\tilde{\rho}^2$ reduced density matrix associated to the replicated fully {\it disconnected} manifold.}\label{rho2_1}
\end{figure}

To compute $\tilde{\rho}^2[1',2' ; 1,2]$, we should start with two copies of the manifold in Fig. \ref{dif_density_to_fill_in}, and then fill in the geometry in bewteen. In Fig. \ref{rho2_to_fill_in} we have depicted the two copies of the geometry in Fig. \ref{dif_density_to_fill_in}, where the repeated numbers indicate the boundaries that are identified by the matrix multiplication, namely,
\be
\tilde{\rho}^2[1',2' ; 1,2]\stackrel{!}{=}\sum_{(3,3')}\tilde{\rho}[1',2' ; 3,3']\tilde{\rho}[3,3' ; 1,2]\label{rho2}.
\ee
The symbol $\stackrel{!}{=}$, in \eqref{rho2} indicates that the matrix $\tilde{\rho}^2$, in Fig. \ref{rho2_to_fill_in} has not been fully specified yet. To fully specify $\tilde{\rho}^2$, we must fill in the geometry in between, and then take a trace over the red segments. At this point is where the ambiguity shows up. There are several ways in which we can fill in the geometry. The first and obvious case is represented in Fig. \ref{rho2_1}

\begin{figure}[h]
\centering
\includegraphics[width=.8\textwidth]{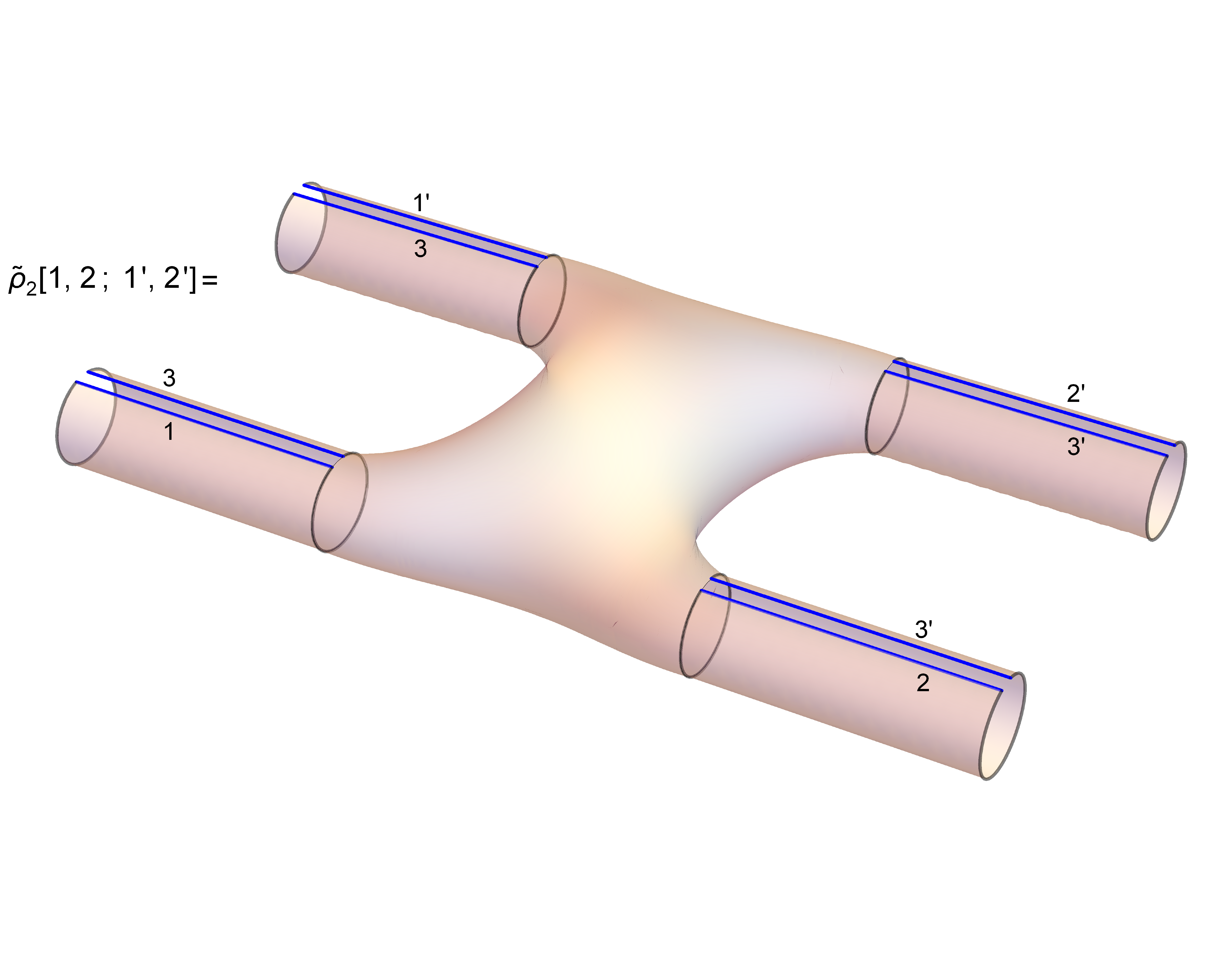}
\caption{\sl Geometric representation of the $\tilde{\rho}_2$ reduced density matrix associated with the replicated {\it connected} manifold. This geometry can be regarded as a complex wormhole connecting the copies.}\label{rho_connected}
\end{figure}

It can be regarded as the genuine $\tilde{\rho}^2[1',2' ; 1,2]$. The word ``genuine'' is in order because $\tilde{\rho}^2[1',2' ; 1,2]$, in Fig. \ref{rho2_1} is the square of the matrix in Fig. \ref{dif_density}. Also, because by filling in the geometry differently we can define another density matrix Fig. \ref{rho_connected}. We shall denote it as $\tilde{\rho}_2[1,2;1',2']$, because it is not the square of the matrix in Fig. \ref{dif_density}.

The geometry is connecting the two copies and can be regarded as a complex wormhole. On the one hand, obviously, it is not the square of the matrix in Fig. \ref{dif_density}. On the other hand, we can see how the ambiguity in extending the $\Sigma_0$ slice has led us to the concept of replica wormhole.

There are more geometries one could include in the density matrix definition. The one we have considered so far in Fig. \ref{rho_connected} is topologically equivalent to an $S^2$, with four punctures. Certainly, we could include the higher genus ones. However, like in the two-dimensional case, we believe that they would be suppressed by some topological mechanism \cite{Penington:2019kki}.

Other connecting-geometries could be considered. For instance, we could connect the two cylinders on the left and the two on the right in Fig. \ref{rho2_to_fill_in}; or the upper cylinder on the left with the lower on the right and the lower on the left with the upper on the right. However, these geometries are not allowed because they do not satisfy the boundary conditions on the red slices. In other words, the red slices can not be fully inscribed in these geometries.

At this point we find it convenient to make a distinction among these matrices. In what follows we regard $\tilde{\rho}(n)$, as the most general density matrix can be associated to a particular, non-fully specified manifold \footnote{By non-fully specified manifold we mean the manifold before having connected all its internal boundaries, as in Fig. \ref{dif_density_to_fill_in}
and Fig. \ref{rho2_to_fill_in}.}, for instance Fig. \ref{rho2_to_fill_in}. To construct
$\tilde{\rho}(n)$, and fully specify it we can procced as in \cite{Page:1986vw}, and in equation \eqref{Tr_mixed_den}. We can consider all the contributions comming from the disconected and connected geometries fulfilling the boundary conditions on the internal boundaries and on the red slices, i.e.,
\be
\tilde{\rho}(n)=\underset{\text{disconnected}}{ \tilde{\rho}^n} +\underset{\text{connected}}{ \tilde{\rho}_n}\label{rho_connected_general}.
\ee
Here, $\tilde{\rho}^n$ is genuinely the {\it n}th power of the matrix $\tilde{\rho}$.

Now, we can regard, for instance, $\tilde{\rho}(2)[1',2' ; 1,2]$, as the most general density matrix we can associate to the configuration depicted in Fig. \ref{rho2_to_fill_in}. Adding all the contributions, it is given by
\be
\tilde{\rho}(2)[1',2' ; 1,2]=\underset{\text{disconnected}}{ \tilde{\rho}^2[1',2' ; 1,2]} +\underset{\text{connected}}{\tilde{\rho}_2[1',2' ; 1,2]}.
\ee
One of the advantage of these distinctions (or definitions) is that we can avoid the factorization problem \cite{Penington:2019kki}. By avoiding this problem, no ensemble average is needed to make the setup consistent.

Having $\tilde{\rho}(n)$, we can compute the following quantity
\be
S=-\lim_{n\rightarrow 1}\partial_{n}\text{Tr}\Big[\tilde{\rho}(n)\Big] = -\lim_{n\rightarrow 1}\ \partial_{n}\text{Tr}\Big[\tilde{\rho}^n +\tilde{\rho}_n\Big].\label{general_entropy}
\ee
This quantity can not be identified as the entanglement entropy of the segments, in the ordinary QFT sense, see \cite{Giddings:2020yes} for a discussion about this idetinfication and other  issues  related to the replica wormhole calculus. The reason is the derivative of $\text{Tr}\Big[\tilde{\rho}(n)\Big]$, does not lead to $-\text{Tr}\Big[\tilde{\rho} \log \tilde{\rho}\Big]$, instead it leads to
\be
S = -\text{Tr}\Big[\tilde{\rho} \log \tilde{\rho}\Big] -\lim_{n\rightarrow 1}\ \partial_n\text{Tr}\Big[\tilde{\rho}_n\Big],
\ee
where the connected contribution appears.
Of course, if we assume that in QG, the definition of entanglement entropy should be generalized to \eqref{general_entropy}, which seems to be supported by \cite{Almheiri:2019qdq, Penington:2019kki, Gautason:2020tmk, Anegawa:2020ezn, Hashimoto:2020cas, Hartman:2020swn}, when using the replica trick, then we would be computing the actual entanglement entropy associated to $\Sigma_{-}$ and $\Sigma_{+}$.

We want to point out the following fact. We have posed the problem of computing the entanglement entropy for two segments on a nice slice for $t_0\neq 0$. Instead, if we had posed the problem for $t_0=0$, where no red slice appears, see Fig. \ref{extension0}, no wormhole would have appeared in the calculation of the entanglement entropy. Of course, after evolving the state in Lorentzian time, we would have room again for including the replica wormholes.

This completes the possing of the problem but a few comments may be in order. At this point one might ask, whether there  are classical solutions to the equations of motion of gravity with matter that allow geometries such that the one depicted in Fig. \ref{rho_connected}, or for higher $n$. In the first place, we should note that  $\text{Tr}\Big[\tilde{\rho}(n)\Big]$  is defined as a path integral, and the path integral does not care about classical solutions. So, at least as off-shell configurations all those geometries will contribute to the density matrix.

In the semiclassical approximation, however, the situation is more delicate. In this approximation the path integral is avaluated on  classical solutions to the equations of motion. Whether or not these solutions exist in four dimensions  is not clear yet. Even for the two dimensional case,  finding solutions on these geometries with non-trivial topologies is a dificult task \cite{Almheiri:2019qdq, Penington:2019kki}.

\section{Conclusions}\label{sec:7}

This paper has combined several ideas to propose a new semiclassical QG  state on a nice slice for a Schwarzschild BH. On these slices, the low energy description remains valid during most of the BH evaporation. For this to happen, a portion of the nice slices inside the BH must be fixed at some $\text{r}_0<2\text{M}$. Because of this fixed portion, the preparation of a semiclassical QG  state by evolution in complex time is not straightforward. The main reason is that the fixed portion does not depend on time explicitly. The only dependence appears on the boundaries of the fixed segment. Moreover, the geometry that describes the semiclassical state's preparation connects the two boundaries of the density matrix, and by no means one can get a disconnected (disjoint) geometry after complex time evolution. Our main result has been to find that the QG state on a nice slice is a global mixed state, similar to \cite{Page:1986vw}.

We have also found, even though the state is not pure, that the thermodynamic entropy associated with the geometry is the expected one for a two-sided BH \eqref{expected_entropy2BH}. For simplicity, in this first proposal, we did not include the matter contribution. We also did not study the time evolution.

After computing the BH thermodynamic entropy, we moved to the entanglement entropy. By possing the problem of computing the entanglement entropy for two segments on a nice slice, we found several new and interesting features.

As we assume that we are performing calculations in QG, we have followed a different logic to that in QFT to build the density matrix associated with the replicated manifold. In QFT, one considers a fixed geometry on the manifold that defines the density matrix associated to some state. The usual replica trick consists in taking {\it n} copies of that manifold,  and  glue them together according to the region we are interested in computing the entanglement entropy. This new manifold defines the reduced density matrix. After extending {\it n} from the Integers to the Reals, we can use it to compute the entanglement entropy according to the usual rules in QFT.

In QG, the fact that neither the geometry nor the topology  are  fixed,  affects the density matrix definition we can associate to the replicated manifold. In fact, it directly affects the very concept of replicated manifold. Also, in QG, there is an exact prescription to prepare a semiclassical QG state through complex time evolution \cite{Hartle:1983ai}. Of course, this prescription is subjected to the appearance of time on those surfaces where we are interested in defining the state.

On the nice slices, the fact that there are portions that do not evolve forward in time  introduces an ambiguity in associating a replicated manifold to a particular density matrix. It has been well illustrated in section \ref{sec:6}. Now, the association is not unique and, to a particular configuration, for instance, in Fig. \ref{rho2_to_fill_in}, we can associate many (perhaps infinitely many, the higher genus geometries) manifolds. In fact adding all possible contributions together would lead to a {\it good} density matrix too, as in \eqref{rho_connected_general}, in the same spirit of \cite{Page:1986vw}. This ambiguity has led us to the concept of replica wormhole connecting different replicas \cite{Almheiri:2019qdq, Penington:2019kki}.

The next step in this construction would be to add matter in it and study the evaporating BH, which is a time-dependent system. The inclusion of matter for dimensions higher than two is not straightforward, mainly because we must consider the backreaction on the metric for the evaporating BH. Although, in principle, one could use the approximation in \cite{Hashimoto:2020cas}.

A more delicate point when adding matter in this setup would be to define the radiation's information flux properly. Usually, it is defined on $\mathcal{I}^{+}$ in the Penrose diagram. Here, however, we have possed the problem of computing the entanglement entropy on segments that extend from a finite $\text{r}=\text{r}_0<2\text{M}$, or $\text{r}=\text{r}_1>2\text{M}$, to infinity on a region where gravity should be consider quantum,  as in \cite{Gautason:2020tmk}.  Note that the segments do not sit completely on $\mathcal{I}^{+}$, as in \cite{Hartman:2020swn}. The key point to properly address this calculation is to note that we can take the limit  $t_0\rightarrow \infty$. As we have shown, the semiclassical geometry does not dependent on the particular choice of $t_0$. When $t_0\rightarrow \infty$, the segments $\Sigma_{-}$ and $\Sigma_{+}$ sit completely on null infinity, i.e., on $\mathcal{I}^{+}$ in the Penrose diagram, see Fig. \ref{NI_nice_slice}.

In reference \cite{Giddings:2020yes}, some criticism related to the connection of the replica wormhole calculation and the amplitudes computed according to the usual rules of QFT was raised. Here we have presented some arguments that partially answer the questions in \cite{Giddings:2020yes}. For instance, in section \ref{sec:5}, we have presented the density matrix interpretation of the geometry we have built here, together with a prescription on how the amplitudes must be assembled to give rise to the density matrix. In Fig \ref{dens_fact1}, we have presented the building blocks of this density matrix. It turns out that the building blocks are density matrices too. Each of them, in turn, would be constituted by wave functionals.

After finishing, we would like to speculate,   as mencioned in footnote \ref{speculate},  about an intriguing possibility related to the steps we have followed here to define the QG state. Suppose we want to define a global  state on the slice $\text{T}_1=0$. As usual, one might think this state is the one leading to the HH state. To get this state we evolve in complex  Schwarzschild time the portion of the slice $\text{T}_1=0$, with $\text{X}_1\geq0$, from $\tau=0$ to $\tau=8\pi\text{M}$. For this we use only the right wedge, or for instance \eqref{E_coor} with $t_0=0$. Leaving these two boundaries free (a Pacman figure), we can define a (reduced) density matrix associated to the segment $\text{T}_1=0$, $\text{X}_1\geq0$. As it is well known, this state is not pure on this segment, and it leads to the known thermodynamic entropy for a BH  see for instance \cite{Harlow:2014yka}.

An awkward feature of the geometry representing the partition function, after tracing the degrees of freedom over the mouth of the Pacman (the disk, see Fig. \ref{E_state3}) is that the thermal circle is homotopically equivalent to any circle on the disk. This is not what is expected to happen in the statistical interpretation of QFT.

Now suppose we follow similar steps in defining the state on the slice $\text{T}_1=0$, but this time we evolve in complex Schwarzschild time both segments on the left and right wedges, similarly to what we have done on a nice slice. The geometry, in this case, would not be a packman figure. Instead, it would be a double Pacman figure with opposite orientation overlapping each other and sharing a single point at the horizon.  This follows directly from \eqref{K_coor} with $t\rightarrow -\text{i}\tau$. This geometry would be similar to the one we have found here. Hence it would lead to a global mixed state. Moreover, the thermal circle (after tracing the degrees of freedom over the two mouths of the double Pacman) would not be homotopically equivalent to any circle on this geometry due to the shared point.

It raises the question, whether we can define a pure global state for the BH geometry on any slice. Notice, for instance, that even at  $\mathcal{I}^{-}$, i.e., at $t_0\rightarrow-\infty$, the state would not be a global pure state.   If our speculations turn out to be correct and we can extend it to other foliations, for instance, the ordinary one for a Schwarzschild BH \eqref{K_coor}, it would have repercussions on the information paradox because of the intrinsic impossibility for defining pure global states.  It means that we do not have to worry about pure states evolving into mixed states in QG.  These repercussions will be studied elsewhere.

\vspace{1cm}

\section*{Acknowledgments}

We are grateful to A. Banerjee, E. Ó. Colgáin, H. Casini, T. Hartman,  J. M. Maldacena and Kanghoon Lee for discussions, useful comments and suggestions. We would especially like to thank the members of APCTP for stimulating questions and discussions during the presentation of this work at APCTP.


\begin{thebibliography}{99}

\bibitem{Lowe:1995ac}
D.~A.~Lowe, J.~Polchinski, L.~Susskind, L.~Thorlacius and J.~Uglum,
``Black hole complementarity versus locality,''
Phys. Rev. D \textbf{52}, 6997-7010 (1995).

\bibitem{Mathur:2009hf}
S.~D.~Mathur,
``The Information paradox: A Pedagogical introduction,''
Class. Quant. Grav. \textbf{26}, 224001 (2009)

\bibitem{Polchinski:2016hrw}
J.~Polchinski,
``The Black Hole Information Problem,'' [arXiv:1609.04036 [hep-th]].

\bibitem{Hawking:1974sw}
S.~W.~Hawking,
``Particle Creation by Black Holes,''
Commun. Math. Phys. \textbf{43}, 199-220 (1975).

\bibitem{Hawking:1976ra}
S.~W.~Hawking,
``Breakdown of Predictability in Gravitational Collapse,''
Phys. Rev. D \textbf{14}, 2460-2473 (1976).

\bibitem{Hartle:1983ai}
J.~B.~Hartle and S.~W.~Hawking,
``Wave Function of the Universe,''
Adv. Ser. Astrophys. Cosmol. \textbf{3}, 174-189 (1987), PhysRevD.28.2960.

\bibitem{Giddings:2020dpb}
S.~B.~Giddings,
``Schr\"odinger evolution of the Hawking state,''
[arXiv:2006.10834 [hep-th]].

\bibitem{Giddings:2017mym}
S.~B.~Giddings,
``Nonviolent unitarization: basic postulates to soft quantum structure of black holes,''
JHEP \textbf{12}, 047 (2017).

\bibitem{Giddings:2012bm}
S.~B.~Giddings,
``Black holes, quantum information, and unitary evolution,''
Phys. Rev. D \textbf{85}, 124063 (2012).

\bibitem{Giddings:2007ie}
S.~B.~Giddings,
``Quantization in black hole backgrounds,''
Phys. Rev. D \textbf{76}, 064027 (2007).

\bibitem{Giddings:2006be}
S.~B.~Giddings,
``(Non)perturbative gravity, nonlocality, and nice slices,''
Phys. Rev. D \textbf{74}, 106009 (2006).

\bibitem{Hawking:1983hj}
S.~W.~Hawking,
``The Quantum State of the Universe,''
Adv. Ser. Astrophys. Cosmol. \textbf{3}, 236-255 (1987).

\bibitem{Gibbons:1976ue}
G.~W.~Gibbons and S.~W.~Hawking,
``Action Integrals and Partition Functions in Quantum Gravity,''
Phys. Rev. D \textbf{15}, 2752-2756 (1977).

\bibitem{Hawking:1978jz}
S.~W.~Hawking,
``Quantum Gravity and Path Integrals,''
Phys. Rev. D \textbf{18}, 1747-1753 (1978).

\bibitem{Hawking:1986vj}
S.~W.~Hawking,
``The Density Matrix of the Universe,''
Phys. Scripta T \textbf{15}, 151 (1987).

\bibitem{Page:1986vw}
D.~N.~Page,
``Density Matrix of the Universe,''
Phys. Rev. D \textbf{34}, 2267 (1986).

\bibitem{Hartman:2013qma}
T.~Hartman and J.~Maldacena,
``Time Evolution of Entanglement Entropy from Black Hole Interiors,''
JHEP \textbf{05}, 014 (2013).

\bibitem{Page:1993wv}
D.~N.~Page,
``Information in black hole radiation,''
Phys. Rev. Lett. \textbf{71}, 3743-3746 (1993).

\bibitem{Page:2013dx}
D.~N.~Page,
``Time Dependence of Hawking Radiation Entropy,''
JCAP \textbf{09}, 028 (2013).

\bibitem{Almheiri:2019qdq}
A.~Almheiri, T.~Hartman, J.~Maldacena, E.~Shaghoulian and A.~Tajdini,
``Replica Wormholes and the Entropy of Hawking Radiation,''
JHEP \textbf{05}, 013 (2020).

\bibitem{Penington:2019kki}
G.~Penington, S.~H.~Shenker, D.~Stanford and Z.~Yang,
``Replica wormholes and the black hole interior,''
[arXiv:1911.11977 [hep-th]].


\bibitem{Gautason:2020tmk}
 F.~F.~Gautason, L.~Schneiderbauer, W.~Sybesma and L.~Thorlacius,
``Page Curve for an Evaporating Black Hole,''
JHEP \textbf{05}, 091 (2020).

\bibitem{Anegawa:2020ezn}
T.~Anegawa and N.~Iizuka,
``Notes on islands in asymptotically flat 2d dilaton black holes,''
JHEP \textbf{07}, 036 (2020).

\bibitem{Hashimoto:2020cas}
K.~Hashimoto, N.~Iizuka and Y.~Matsuo,
``Islands in Schwarzschild black holes,''
JHEP \textbf{06}, 085 (2020).

\bibitem{Hartman:2020swn}
T.~Hartman, E.~Shaghoulian and A.~Strominger,
``Islands in Asymptotically Flat 2D Gravity,''
JHEP \textbf{07}, 022 (2020)
doi:10.1007/JHEP07(2020)022.

\bibitem{hartman_lec}
Thomas Hartman, ``Lectures on Quantum Gravity and Black Holes,''
http://www.hartmanhep.net/topics2015/gravity-lectures.pdf.

\bibitem{Anous:2020lka}
 T.~Anous, J.~Kruthoff and R.~Mahajan, ``Density matrices in quantum gravity,''
SciPost Phys. \textbf{9}, no.4, 045 (2020), 
[arXiv:2006.17000 [hep-th]].

\bibitem{Chen:2020tes}
Y.~Chen, V.~Gorbenko and J.~Maldacena,
``Bra-ket wormholes in gravitationally prepared states,''
[arXiv:2007.16091 [hep-th]].

\bibitem{DeWitt:1967yk}
B.~S.~DeWitt,
``Quantum Theory of Gravity. 1. The Canonical Theory,''
Phys. Rev. \textbf{160}, 1113-1148 (1967).

\bibitem{Halliwell:1989dy}
J.~J.~Halliwell and J.~B.~Hartle,
``Integration Contours for the No Boundary Wave Function of the Universe,''
Phys. Rev. D \textbf{41}, 1815 (1990).

\bibitem{Hartle:1976tp}
J.~B.~Hartle and S.~W.~Hawking,
``Path Integral Derivation of Black Hole Radiance,''
Phys. Rev. D \textbf{13}, 2188-2203 (1976).

\bibitem{Deser:1959zza}
R.~Arnowitt and S.~Deser,
``Quantum Theory of Gravitation: General Formulation and Linearized Theory,''
Phys. Rev. \textbf{113}, 745-750 (1959).

\bibitem{Arnowitt:1959ah}
R.~L.~Arnowitt, S.~Deser and C.~W.~Misner,
``Dynamical Structure and Definition of Energy in General Relativity,''
Phys. Rev. \textbf{116}, 1322-1330 (1959).

\bibitem{Arnowitt:1960es}
R.~L.~Arnowitt, S.~Deser and C.~W.~Misner,
``Canonical variables for general relativity,''
Phys. Rev. \textbf{117}, 1595-1602 (1960).

\bibitem{Brown:1990fk}
J.~D.~Brown, E.~A.~Martinez and J.~W.~York, Jr.,
``Complex Kerr-Newman geometry and black hole thermodynamics,''
Phys. Rev. Lett. \textbf{66}, 2281-2284 (1991).

\bibitem{Maldacena:2001kr}
J.~M.~Maldacena,
``Eternal black holes in anti-de Sitter,''
JHEP \textbf{04}, 021 (2003).

\bibitem{Harlow:2014yka}
D.~Harlow,
``Jerusalem Lectures on Black Holes and Quantum Information,''
Rev. Mod. Phys. \textbf{88}, 015002 (2016),
[arXiv:1409.1231 [hep-th]].

\bibitem{Giddings:2020yes}
S.~B.~Giddings and G.~J.~Turiaci,
``Wormhole calculus, replicas, and entropies,''
[arXiv:2004.02900 [hep-th]].

\end{thebibliography}
\end{document}